\documentclass[12pt]{iopart}

\pdfoutput=1
\usepackage{amsfonts}
\usepackage{graphicx}
\usepackage{hyperref}
\usepackage{color}
\usepackage{subfigure}
\usepackage{textcomp}
\usepackage{cite}
\usepackage{float}
\usepackage{soul}
\usepackage{stackrel}
\usepackage{tikz}
\usepackage[mathscr]{eucal}
\graphicspath{{figures/}}

\newenvironment{rcases}
  {\left.\begin{aligned}}
  {\end{aligned}\right\rbrace}

\definecolor{dgreen}{rgb}{0,0.7,0}

\expandafter\let\csname equation*\endcsname\relax
\expandafter\let\csname endequation*\endcsname\relax
\usepackage{amsmath,amssymb}

\pdfoutput=1
\usepackage{esint}

\usepackage{color}
\definecolor{dgreen}{rgb}{0,0.7,0}

\newcommand{\beq}{\begin{equation}}
\newcommand{\eeq}{\end{equation}}
\newcommand{\bea}{\begin{eqnarray}}
\newcommand{\eea}{\end{eqnarray}}

\begin{document}

\title[]{Tracer dynamics in the active random average process}

\author{Saikat Santra$^1$, Prashant Singh$^{2}$ and Anupam Kundu$^1$}

\address{International Centre for Theoretical Sciences, Tata Institute of Fundamental ~Research, Bengaluru 560089, India $^1$}
\address{Niels Bohr International Academy, Niels Bohr Institute,
University of Copenhagen, Blegdamsvej 17, 2100 Copenhagen, Denmark $^2$}
\ead{prashant.singh@nbi.ku.dk, saikat.santra@icts.res.in}
\vspace{10pt}

\begin{abstract}
We investigate the dynamics of tracer particles in the random average process (RAP), a single-file system in one dimension. In addition to the position, every particle possesses an internal spin variable $\sigma (t)$ that can alternate between two values, $\pm 1$, at a constant rate $\gamma$. Physically, the value of $\sigma (t)$ dictates the direction of motion of the corresponding particle and for finite $\gamma$, every particle performs a non-Markovian active dynamics. Herein, we study the effect of this non-Markovianity in the fluctuations and correlations of the positions of tracer particles. We analytically show that the variance of the position of a tagged particle grows sub-diffusively as $\sim \zeta_{\text{q}} \sqrt{t}$ at large times for the quenched uniform initial condition. While this sub-diffusive growth is identical to that of the Markovian/non-persistent RAP, the coefficient $\zeta_{\text{q}} $  is rather different and bears the signature of the persistent motion of active particles through higher point correlations (unlike in the Markovian case). Similarly, for the annealed (steady state) initial condition, we find that the variance scales as $\sim \zeta_{\text{a}} \sqrt{t}$ at large times with coefficient $\zeta_{\text{a}} $ once again different from the non-persistent case. Although $\zeta_{\text{q}}$ and $\zeta_{\text{a}} $ both individually depart from their Markovian counterparts, their ratio $\zeta_{\text{a}} / \zeta_{\text{q}}$ is still equal to $\sqrt{2}$, a condition observed for other diffusive single-file systems. This condition turns out to be true even in the strongly active regimes as corroborated by extensive simulations and calculations. Finally, we study the correlation between the positions of two tagged particles in both quenched uniform and annealed initial conditions. We verify all our analytic results by extensive numerical simulations.
\end{abstract}

\section{Introduction}
The dynamics of a tracer particle in a collection of non-overtaking particles in one dimension is a prototypical example of strongly correlated system in statistical physics. This non-overtaking constraint, called single-file constraint, drastically changes the dynamical behaviour of a tracer particle \cite{Harris1965, Jepsen1965, Benichou2018, Arratia1983,Alexander1978, Rodenbeck1998, Barkai2009, Kravpivsky2014, Krapivsky2015, Tirtha2022, Abhishek-Sanjib}. For example, in single-file diffusion, the mean squared displacement (MSD) of a tagged particle grows sub-diffusively as $\sim \sqrt{t}$ at late times in contrast to the linear growth of the MSD of a free diffusive particle. The coefficient of the sub-diffusive growth depends on the particle number density and bare diffusion of the particles\cite{Jepsen1965, Alexander1978, Arratia1983}. For Hamiltonian systems,  the single-file constraint, slows down the motion of a tagged particle \cite{Abhishek-Sanjib, Anjan, Kollman-PRL-2003, Cividini2016}. This slowing down of the dynamics is a common effect in the single-file motion and occurs due to the hindrance in the motion faced by one particle due to the presence of other particles. It has been found that the coefficient of the late time growth of the MSD of a tagged particle crucially depends, in addition to the density of the particles, on the microscopic dynamics of individual particles, interactions among themselves and on the statistical properties of the initial state \cite{Kollman-PRL-2003, Anjan, Cividini-2017, elibarkai, Krapivsky2015, Tirtha2022}. In this paper, we study how the motion of a tagged particle gets modified if all particles in the single-file system are active.

Active matter is a class of driven out-of-equilibrium systems where every individual unit consumes energy from the environment and converts it into a systematic movement via some internal mechanisms \cite{Ramaswamy2010, Marchetti2013}. At the collective level, these particles exhibit interesting phenomena such as motility-induced phase separation, absence of equation of state for pressure etc \cite{Cates2015, Solon2015, Soto_PRE_2014,Soto_PRE_2016}. Non-interacting active particles also show behaviours which are different than their passive counterpart as exemplified by clustering inside bounded domain, climbing against potential hill, non-Boltzmann stationary state and large deviation in position distributions and survival probability which are different than the thermal particles \cite{dist-1, Urna, dist-2, dist-3,dist-4, dist-5, new-dist-1, FPT-1, Mori,FPT-3, FPT-4, FPT-5, FPT-6, local, convex-1, convex-2,Woillez2019, Singh2022,Banerjee2020}. At the interacting front, the distribution of two active particles with mutual exclusion has been studied and shown to display jamming features \cite{Slowman2016}. Going beyond two particles, there have also been attempts to derive fluctuating hydrodynamic descriptions for active lattice gases that turn out to be useful to study density and current fluctuations and entropy production \cite{Jose2023, FHT-1, FHT-2}. With the growing interest in active matter in the last decade, people naturally got interested in knowing how broken detailed balance at the microscopic dynamics manifests itself in the tracer dynamics in single-file motion. Several numerical studies in this direction have pointed out that the temporal growth of the MSD of a tagged particle, at late times, remains same as in the absence of activity. However, the coefficient associated with this temporal growth gets non-trivially modified due to the persistent nature of these particles \cite{Teomy2019-1, Teomy2019-2, Galanti2013, Kabir, Pritha, Tirthankar,Put2019}. Attempts to derive this with harmonic chain of active particles reproduce only the passive result at late times and do not shed light on the role of activity in the tracer dynamics \cite{Put2019,prashant-harm-29-34}. It is, however, not difficult to realize that the presence of activity increases the correlations among particles and this, in addition to the single-file constraint, should affect the motion of a tagged particle. Naturally one may ask how such enhanced correlations affect the motion of a tagged particle? While the above mentioned studies discuss the overall effect of the presence of activity, the contribution from the enhanced correlation is not very clear and transparent. Moreover, how the fluctuations in the initial conditions affect the motion of tracer particles in active single-file systems is also not explored. In absence of a general formulation to investigate these questions, it is crucial to study specific model systems that are amenable to analytical calculations. In this paper, we consider a version of the random average process (RAP) in which individual particles are subjected to active noises. For this model, we provide systematic answers to these questions.

Originally, the RAP model was first studied for non-persistent particles (without active noises) by Fontes and Ferrari as a generalization of the smoothing process and the voter model \cite{Ferrari1998}. It has also appeared in several other physical problems like force fluctuations in bead packs \cite{Coppersmith1996}, in mass transport models \cite{Rajesh2000, Krug2000}, in models of wealth distribution and traffic \cite{Ispolatov, Mezak} and generalized Hammersley process \cite{Aldous}. In this model, motion of the particles is restricted as each particle can jump, with a fixed rate $p$, on either side only by a random fraction $\eta$ of the space available till the next neighbouring particle \cite{Ferrari1998, Rajesh2001}. The random fraction $\eta$ is chosen from some distribution $R(\eta)$. Since particles cannot overtake their neighbouring particles, their initial order remains preserved throughout the time evolution which gives rise to the single-file motion. Meanwhile, the jump that a particle makes at a given instant is independent of what it does in the previous step. Therefore, we will refer to this model as the Markovian RAP (MRAP). Later, we will contrast this with the active case where every particle possesses a spin variable $\sigma(t)$ that dictates the direction of its motion and has non-vanishing correlations at two different times.

Being a paradigmatic model for interacting multi-particle system, the motion of tagged particles in MRAP was studied previously and several results were obtained both analytically and numerically.  If $x_i(t)$ denotes the position of the $i$-th particle at time $t$, then the MSD $ \langle z_i ^2(t) \rangle$ (with $z_i(t) = x_i(t)-x_i(0)$) and the correlation between two tagged particles $ \langle z_i (t) z_j(t) \rangle$ were computed using microscopic calculations \cite{Rajesh2001, Krug2000, Schutz2000, CividiniKundu2016}, as well as using hydrodynamic approach \cite{JCividini2016}. At late times, these quantities are explicitly given by
\begin{equation}
\begin{rcases}
& \langle z_0 ^2(t) \rangle   \simeq \frac{\rho^{-2} \mu _2 \sqrt{\mu _1 p} }{(\mu _1- \mu _2) \sqrt{\pi}}~\sqrt{t},\\
& \langle z_0 (t) z_i(t) \rangle   \simeq \frac{\rho^{-2} \mu _2 \sqrt{\mu _1 p}} {(\mu _1- \mu _2) \sqrt{\pi}}~\sqrt{t}~f \left( \frac{|i|}{\sqrt{4 \mu _1 p t}}\right),
\end{rcases}
\text{ (MRAP)} 
\label{passive-case}
\end{equation}
where $\mu_i$ is the $i^{\rm th}$ moment of the jump distribution $R(\eta)$ and $\rho$ is the stationary density of the particles. Explicit expression of the scaling function $f(y)$ is given by 
\cite{Rajesh2001, JCividini2016}
\begin{align}
f(y) = e^{-y^2}-\sqrt{\pi} y ~\text{erfc}(y). \label{scaling-func}
\end{align}
The same scaling function also appears in many single-file systems that possess diffusive hydrodynamics at the macroscopic scales \cite{JCividini2016, punyabrata-kundu, Benichou2018, Benichou-1, Benichou-2}. It is, however, important to note that for MRAP, the equations for two-point correlation functions close onto themselves. Therefore, two-point correlations are enough to decide the pre-factor in the expressions of MSD. Contrarily, this is found not to be true when the particles are subjected to active noises. Then, the two-point correlations would depend on three-point correlations and so on. Consequently, higher point correlations start to contribute to the growth of the MSD through two-point correlations. For example, such dependence on higher order correlations was also observed for gap statistics in hardcore run and tumble particles \cite{Rahul-Rajesh}. Question then arises: does activity facilitate in the growth of MSD? If so, how? In this paper, we present an example of a model where these questions can be thoroughly addressed through microscopic analytic computations aided by numerical simulations.

Our paper is organised as follows: In Section \ref{model}, we introduce the model, fix notations and also summarize main results of the paper. We then compute the spin-position correlation function $C_i(t) = \langle z_i(t) \sigma_0(t)\rangle $ in Section \ref{corr-ci}. We use this result to study the properties of tracer particles with fixed initial condition in Section \ref{sec-q}. More specifically, we look at the MSD of the position of a tagged particle in Section \ref{msd-q} and position correlation of two tagged particles in Section \ref{correlation-q}. Section \ref{sec-a} discusses these quantities in the annealed case with Section \ref{msd-a} devoted to the MSD and Section \ref{correlation-a} to the correlation function. Since most of our analytic results are derived with mean-field approximation, we discuss the validity of our results for strongly active regime (small $\gamma$) in Section \ref{small-gamma}. Finally we conclude in Section \ref{conclusion}.

\begin{figure}[]
	\includegraphics[scale=0.5]{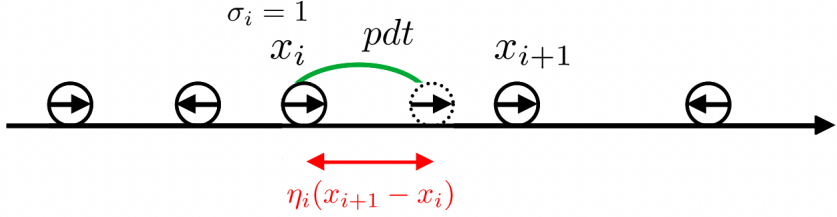}
	\centering
	\caption{Schematic illustration of the random average process with active particles. Each particle has an internal spin denoted by the arrows inside the circles and the direction of the arrow represents the state of the spin. The spin variable of individual particles changes direction independently with rate $\gamma$. In a small time interval $dt$ a particle chosen at random either makes a jump in the direction of the spin with probability $pdt $ or does not jump with the remaining probability $(1-pdt)$. Every successful jump of a particle, say $i^{\rm th}$ with $\sigma_i=1$,  takes place by a random fraction $\eta_i$ of the space available in jump direction {\it i.e.} by an amount $\eta _i \left(x_{i+1}(t)-x_i(t) \right)$ where $\eta _i \in [0,1)$ is a random variable drawn from the distribution $R(\eta _i)$.}    
	\label{scheme}
\end{figure}
\section{Model and summary of our main results}
\label{model}
We consider active particles moving in an infinite line distributed with density $\rho$. We denote the position of the $i$-th particle at time $t$ by $x_{i}(t)$ where $i \in \mathbb{Z}$ and $x_i(t) \in \mathbb{R}$. In addition, every particle has an internal variable $\sigma _i(t)$ (called spin) which can alternate between $\pm 1$ at a rate $\gamma>0$. The variable $\sigma _i(t)$ represents the usual dichotomous noise widely studied for the run and tumble particles \cite{dist-1}. Initially, the positions of these particles are fixed and are kept at a fixed distance $a = 1/\rho$ apart. However, the initial value of $\sigma _i(0)$ can be random which, for simplicity, we choose to be $\pm 1$ with equal probability $1/2$. Thus for all $i \in \mathbb{Z}$, we have
\begin{align}
x_i(0) &= i a = i / \rho, \label{quenched-eq}\\
\sigma _i(0) &= \begin{cases}
   ~~1~~\text{with probability }1/2, \\ 
   -1~~\text{with probability }1/2.
\end{cases}
\end{align}
At a small time interval $[t, t+dt]$, the direction of motion of the $i$-th particle depends on its spin variable $\sigma _i(t)$. If $\sigma _i(t) = 1$, then the particle jumps to the right with probability $p dt$ and does not jump with probability $(1-pdt)$. On the other hand, if $\sigma _i(t)=-1$, then the particle jumps to its left with probability $p dt$ and with probability $(1-pdt)$, stays at $x_i(t)$.  The jump, either to the left or to the right, is by a random fraction $\eta_i$ of the space available between the particle and its neighbour. This means, when the particle jumps to the right, it will jump by an amount $\eta_i \left[x_{i+1}(t)-x_i(t) \right]$ whereas jump to the left takes place by an amount $\eta_i \left[x_{i-1}(t)-x_i(t) \right]$. The jump fraction $\eta _i \in [0,1)$ is a random variable drawn independently from the distribution $R(\eta)$ and characterized by the moments $\mu _k = \langle \eta ^k \rangle$. In contrary to the original MRAP, we see that the motion of a particle at time interval $[t, t+dt]$ depends on the value of $\sigma (t)$ which itself depends on its previous history. Thus, every particle performs a non-Markovian active dynamics. A schematic illustration of this model is shown in Figure \eqref{scheme}. We refer to this model as the active random average process (ARAP).

\noindent
The time evolution equation for the position $x_i(t)$ and spin $\sigma_i(t)$ can be written as
\begin{align}
&x_i(t+ d t)=x_i(t)+\Gamma_i(t),  \label{update-eq-1}\\
&\sigma_i(t+dt)=\begin{cases}-\sigma_i(t),~~~\text{with probability} ~~~~~\gamma dt, \\
~~\sigma_i(t), ~~~\text{with probability} ~~(1-\gamma dt),
\end{cases} 
\label{update-eq-2}
\end{align}
where the increment $\Gamma _i(t)$ reads
\bea
\label{eq:eta}
\Gamma_i(t)=\begin{cases}
	\eta_i \left[x_{i+1}(t)-x_i(t) \right],~~~\text{with probability}~~~~ \left(\frac{1+\sigma_i(t)}{2} \right) pdt ,\\
	\eta_i \left[x_{i-1}(t)-x_i(t) \right],~~~\text{with probability}~~~~ \left(\frac{1-\sigma_i(t)}{2} \right) pdt ,\\
	~~~~~~~0, ~~~~~~~~~~~~~~~~~~\text{with probability}~~~~~~~ (1-pdt).
\end{cases}	
\eea
Since we are interested in studying the variance and correlation of the displacement of the tagged particles, it seems convenient to work in terms of the displacement variable $z_i(t)=x_i(t)-x_i(0)$. Also, the rate $p$, essentially fixes a time scale in the model and can be scaled out by redefining $\gamma \to \gamma/p$. Hence, from now on, we choose $p=1$ without any loss of generality. The aforementioned update rules then become 
\begin{align}
& z_i(t+dt) =z_i(t)+\Gamma_i(t), ~~~\text{with }\label{update-eq-3} \\
& \Gamma_i(t)= \begin{cases}
\eta_i \left[z_{i+1}(t)-z_i(t) +a\right],~~~\text{with probability}~~~~ \left(\frac{1+\sigma_i(t)}{2} \right) dt ,\\
\eta_i \left[z_{i-1}(t)-z_i(t)-a \right],~~~\text{with probability}~~~~ \left(\frac{1-\sigma_i(t)}{2} \right) dt ,\\
~~~~~~~~~~0,~~~~~~~~~~~~~~~~~~~~\text{with probability}~~~~~~~ (1-dt).
\end{cases}
\label{update-eq-4}	
\end{align}
For the Markov case $(\gamma \to \infty)$, we saw in Eq. \eqref{passive-case} that the variance and the equal time correlation function exhibit interesting scaling behaviours with time. Here, we derive them for finite $\gamma$ illustrating the interplay of single-file constraint and non-Markovianity in the dynamics. Before, delving into the main calculation, we present a brief summary of all our results in the paper.
\subsection{Summary of the main results}

\begin{itemize}
	\item We first study the two-point correlation  $C_i(t ) \equiv \langle z_0(t) \sigma _i(t) \rangle$ between the displacement of one particle and the spin of another particle. This basic quantity turns out to be instrumental in determining the two-point position correlations for the active RAP. However, it is absent in systems without persistent noise $\sigma_i(t)$ {\it i.e.} in the Markovian RAP system \cite{Rajesh2001}. In Section \ref{corr-ci}, we obtain an explicit expression of $C_i(t)$ based on mean-field (MF) approximation (see Eq.~\eqref{soln_ci}). We observe that this correlation, starting from zero, increases with increasing $t$ (see Figure \eqref{fig-1}) and saturates to a constant value as $t \to \infty $. This value is given by
	\begin{align}
	& C_i(t\to \infty)=\frac{\mu_1 a }{2 \sqrt{ \gamma^2+\mu_1  \gamma}} ~\text{exp}{[-|i|/\xi]}, ~~~~~\text{with } \label{ci_steady_state0} \\ 
	&  \xi^{-1}=\log\left[ \frac{2\gamma+\mu_1 +2\sqrt{ \gamma^2+\mu_1 \gamma}}{\mu_1 } \right].
	\label{ci_steady_state-10}
	\end{align}
	We find in Figure \eqref{fig-2} that this analytical form shows a good agreement with numerical simulations except for small $\gamma$.
	
	\item 
	We next study the temporal behaviour of the MSD $\langle z_0^2(t)\rangle$ of a tagged particle. Similar to MRAP, we find a crossover of $\langle z_0^2(t)\rangle$ from a linear growth at small times to a sub-diffusive growth $(\sim \sqrt{t})$ at large times as
	\begin{equation}
	\langle z_0^2(t)\rangle \simeq \begin{cases}
	\mu_2 a^2  t,  ~~~~~~~\text{for small}~t \\
	\zeta_1 \sqrt{t} +\zeta_2, ~~\text{for large}~t.
	\end{cases}
	\label{eq_MSD}
	\end{equation}
	However, in contrast to the Markov case, constants $\zeta _1$ and $\zeta _2$ are found to depend on the higher-order correlation functions for the ARAP as
	\begin{align}
	& \zeta _1 = \sqrt{\frac{\mu_1}{\pi}} \left[  2a C_I +  T_I + \frac{a \mu _2 }{\mu_1-\mu_2}        \left\{ a + 2 C_1(t \to \infty ) \right\}   \right], \label{zetaeq-1} \\
	&\text{where}~~C_I=\sum _{i=-\infty}^{\infty} C_i(t\to \infty),~~~T_I=\sum _{i=-\infty}^{\infty}  T_i(t \to \infty), \label{def-C_I-T_I}\\
	&~~~~~~~\text{and}~~ \zeta _2 = -\frac{\mu _2 \zeta _1}{4(\mu _1 - \mu _2)}~\sqrt{\frac{\pi}{\mu _1}} .
	\label{zetaeq-2}
	\end{align}
	Here $C_I$ and $T_I$ are constants that depend on the large time saturation values of the two-point correlation $C_i(t)$ and the three-point correlation $T_i(t)$ defined in Eq.\eqref{T-exp}. In the Markovian limit $(\gamma \to \infty)$, both of them vanish and we recover Eq. \eqref{passive-case} for $\langle z_0^2(t)\rangle$. However, for finite $\gamma$, these higher order correlations are non-vanishing and we observe a substantial enhancement of the MSD in comparison to the non-persistent case (see Figure \eqref{fig-MSD}).
	
	\item 
	{We mentioned earlier that the hierarchy of the correlation functions does not close. This is also seen in the time evolution equation of the three-point correlations $T_i(t)$ which reveals that they depend on the four-point correlations which in turn, depend on the higher-point correlations. We numerically demonstrate that any decoupling approximation to break this hierarchy such as decomposing the four-point functions into lower point correlation functions does not provide a good approximation (see \ref{MF_four_point}).}
		
	\item We also compute the position correlation $g_i(t) = \langle z_0(t) z_i(t) \rangle $ and find the following scaling behaviour at large times:
	\begin{align}
	g_i(t)  \simeq   \zeta _{1} \sqrt{t}~ f \left( \frac{|i|}{\sqrt{4 \mu_1 t}}\right), \label{corr}
	\end{align}
	where $f(y)$ is the same scaling function given in Eq. \eqref{scaling-func}. Once again we notice that while the scaling function $f(y)$ is same as the Markov case \cite{Rajesh2001}, the pre-factor $\zeta _1$ in Eq. \eqref{corr} is different and therefore, carries the effect of the persistent dynamics of the active particles.
	
	\item The previous results are derived for the quenched uniform initial condition where the initial positions of the particles remain fixed for different realisations. We also investigate the variance and the correlations with the steady state initial condition. To achieve this, we first evolve the system till time $t_0$ and then start measuring the position till further time $(t_0 +t)$. Observe that the position of the particle at the onset of the measurement is different for different realisations. Taking $t_0 \to \infty$, we obtain the MSD and correlation in the steady state to be 
	\begin{align}
	l_0(t)& = \lim _{t_0 \to \infty} \langle \left[  x_0(t_0+t)-x_0(t_0) \right]^2\rangle \simeq \zeta _1 \sqrt{2 t}, \label{anneal-var-2} \\
	l_i(t)& = \lim _{t_0 \to \infty} \langle \left[  x_i(t_0+t)-x_i(t_0) \right] \left[  x_0(t_0+t)-x_0(t_0) \right]\rangle, \nonumber \\
	&\simeq \zeta _1 \sqrt{2 t} ~f \left( \frac{|i|}{\sqrt{2 \mu _1  t}}\right).
	\end{align}
	Both these results are valid only at large times. Once again, compared to the MRAP, the persistent nature only affects the coefficient $\zeta _1$ but does not change the sub-diffusive exponent. Another interesting observation is that for ARAP also, the ratio of the MSDs in the annealed and quenched initial settings $l_0(t)/g_0(t)$ is equal to $\sqrt{2}$ at large times, a condition valid for the Markov case \cite{Rajesh2001}. This means while both $l_0(t)$ and $g_0(t)$ individually change due to the persistent dynamics, their ratio is still fixed to the value $\sqrt{2}$, same as the Markov case, even at finite $\gamma$.
\end{itemize}

\section{Correlation $C_i(t) = \langle z_i(t) \sigma_0(t)\rangle $}
\label{corr-ci}
Let us begin by computing the spin-position correlation function $C_i(t) = \langle z_i(t) \sigma_0(t)\rangle $ which will be useful in calculating the correlations and fluctuations of the tagged particles later. First note that due to the symmetry of our model, one also has $C_i(t) = \langle z_0(t) \sigma_i(t)\rangle $. In order to evaluate the evolution of $C_i(t)$ in a small time interval $dt$, let us look at the different contributions to $C_i(t+dt) = \langle z_i(t+dt) \sigma_0(t+dt)\rangle $. From this expression, it is clear that the change in $C_i(t+dt)$ can occur either due to the change in $z_i(t+dt)$ or in $\sigma_0(t+dt)$. Following Eq.~\eqref{update-eq-2}, we know that for a small $dt$, $\sigma_0(t+dt) = -\sigma _0(t)$ with probability $\gamma dt$ while $\sigma_0(t+dt) = \sigma _0(t)$ with the complementary probability $(1-\gamma dt)$. This gives rise to the following change in $C_i(t+dt)$:
\begin{align}
C_i(t+dt) &= -\gamma dt ~ \langle z_i(t) \sigma_0(t)\rangle+(1-\gamma dt) ~ \langle z_i(t+d t) \sigma_0(t)\rangle.
\label{eq:C_it_expansion}
\end{align}
Next, we plug the expression of $z_i(t+dt)$ from Eq.~\eqref{update-eq-3} and retain all terms up to linear order in $dt$. It turns out that one then gets different contributions in $C_i(t+dt)$ depending on whether $i=0$ or $i \neq 0$. In particular, for the latter case, we obtain
\begin{align}
C_i(t+dt) & ~=~ C_i(t)-2 \gamma C_i(t) dt +\frac{\mu_1  dt}{2} \left[C_{i+1}(t)+C_{i-1}(t)-2 C_i(t) \right] \nonumber \\
& ~~~~~~~~~~~ + \frac{\mu_1  d t}{2} \left[ \langle \sigma_0(t) \sigma_i (t) z_{i+1} (t)  \rangle -\langle \sigma_0(t) \sigma_i(t) z_{i-1}(t) \rangle \right], \label{time-cii}
\end{align}
where $\mu _k = \langle \eta ^k \rangle = \int _0 ^{1} \eta ^k R(\eta) d \eta$. On the other hand for $i=0$, similar treatment yields
\begin{align}
C_0(t+dt) & ~=~ C_0(t)-2 \gamma C_0(t) dt +\frac{\mu_1  dt}{2} \left[C_{1}(t)+C_{-1}(t)-2 C_0(t)+2a\right].
\label{time-ci0}
\end{align}
Now by combining Eqs.~\eqref{time-cii} and \eqref{time-ci0}, one can appropriately write $C_i(t+dt)$ for any value of $i$ as
\begin{align}
C_i(t+dt) & ~=~ C_i(t)-2 \gamma C_i(t) dt +\frac{\mu_1  dt}{2} \left[C_{i+1}(t)+C_{i-1}(t)-2 C_i(t) +2a \delta_{i,0}\right] \nonumber \\
& ~~~~~~~~~~~ + \frac{\mu_1  d t}{2} \left[ \langle \sigma_0(t) \sigma_i (t) z_{i+1} (t)  \rangle -\langle \sigma_0(t) \sigma_i(t) z_{i-1}(t) \rangle \right]. \label{time-ci}
\end{align}
Taking $dt \to 0$ limit, one arrives at the following equation for $C_i(t)$:
\begin{align}
\frac{dC_i(t)}{dt}&=-2\gamma C_i(t)+ \frac{\mu_1 }{2} \left[ C_{i+1}(t)+C_{i-1}(t)-2 C_i(t)+2 a \delta_{i,0} \right]  \nonumber \\
& ~~~~~~~~+ \frac{\mu_1 }{2} \left[  \langle \sigma_0 (t) \sigma_i(t) z_{i+1}(t)\rangle -\langle \sigma_0 (t) \sigma_i (t) z_{i-1} (t) \rangle \right]. \label{ceq-1}
\end{align}

\begin{figure}[t]
	\includegraphics[scale=1.2]{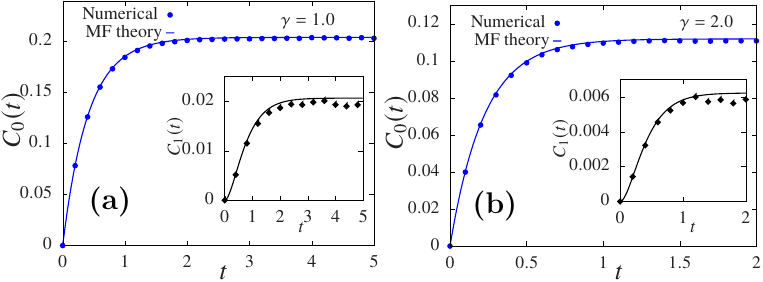}
	\centering
	\caption{Comparison of the analytical (mean-field) expression of the correlation $C_i(t)=\langle z_i(t) \sigma_0(t)\rangle $ in Eq. \eqref{soln_ci} with the numerical simulation for (a) $i=0,~\gamma =1$ (left panel) and (b) $i=0,~\gamma =2$ (right panel). In both panels, insets show the same comparison for $i=1$ with the same set of parameters. Simulation is conducted with $N=201$ particles and the averaging is done over $10^7$ realisations.}    
	\label{fig-1}
\end{figure}
\noindent
While this is an exact time evolution equation, it is not closed due to the presence of higher point correlations. In fact, this turns out to be a general property of the persistent case that the dynamics of any correlation function requires knowledge of higher order correlation functions. This makes the problem analytically challenging. However, as often done, one can make progress by performing the mean field approximation under which we break the three point correlation in Eq. \eqref{ceq-1} into a product of lower order correlations. The validity of this approximation will be discussed later. Proceeding ahead, Eq. \eqref{ceq-1} then simplifies to
\beq
\frac{dC_i(t)}{dt}=-2\gamma C_i(t)+ \frac{\mu_1 }{2} \left[ C_{i+1}(t)+C_{i-1}(t)-2 C_i(t)+2 a \delta_{i,0} \right] . \label{ceq-2}
\eeq 
One needs to solve this equation with the initial condition $C_i(0) = 0$ because $z_i(0)=0$ by definition. In \ref{appen-ceq}, we have explicitly solved it and obtained the expression of $C_i(t)$ as
\beq
C_i(t)=\mu_1 a \int_{0}^{t} e^{-(2\gamma +\mu_1 ) \tau }~ I_{|i|}(\mu_1  \tau) ~d \tau. 
\label{soln_ci}
\eeq
where $I_{i}(y)$ denotes the modified Bessel function. We also see $C_i(t) = C_{-i}(t)$ since the dynamics of the $0^{\rm th}$ particle experiences (statistically) same contributions from particles on its either sides. It turns out that for later calculations, one needs to specify $C_i(t \to \infty)$ which can be easily computed from Eq. \eqref{soln_ci}. Once again we refer to \ref{appen-ceq} for details on this calculation and quote only the final result here as
\begin{align}
& C_i(t\to \infty)=\frac{\mu_1 a }{2 \sqrt{ \gamma^2+\mu_1  \gamma}} ~\text{exp}{[-|i|/\xi]}, ~~~~~\text{with } \label{ci_steady_state} \\ 
&  \xi^{-1}=\log\left[ \frac{2\gamma+\mu_1 +2\sqrt{ \gamma^2+\mu_1  \gamma}}{\mu_1 } \right].
\label{ci_steady_state-1}
\end{align}

\begin{figure}[t]
	\includegraphics[scale=1.0]{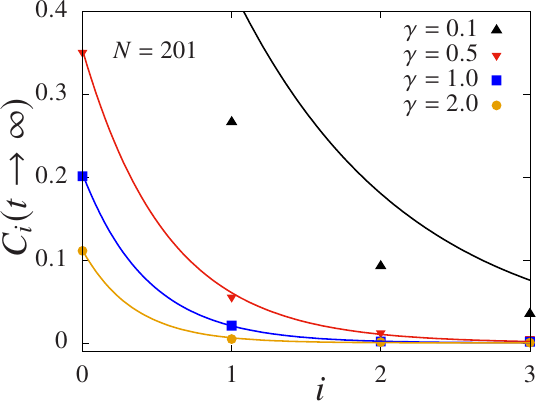}
	\centering
	\caption{Numerical verification of the correlation $C_i(t)=\langle z_i(t) \sigma_0(t)\rangle$ at $t \to \infty$ for different values of $\gamma$. Solid lines represent the analytical (mean-field) expression in Eq. \eqref{ci_steady_state} and symbols are the simulation data. For very small $\gamma$, we observe deviation of the numerical data from the theoretical expression because the mean-field approximation (used in the analytical calculation) becomes less and less valid as $\gamma$ becomes small. The numerical data is obtained using $10^7$ realisations. }
	\label{fig-2}
\end{figure}
\noindent
In Figures \eqref{fig-1} and \eqref{fig-2}, we have compared our analytical results based on the mean field approximations with numerical simulations for different values of $\gamma$. From this comparison, we find that Eq. \eqref{ci_steady_state} matches with the numerics only for moderate and large values of $\gamma$ [see Figure \eqref{fig-2}]. However, for small $\gamma$, our results deviate significantly as seen for $\gamma =0.1$. This is because, at smaller values of $\gamma$, the effect of activity is so strong that the mean field (decoupling) approximation fails and one cannot really neglect the three-point (connected) correlation. We will delve more into this in Section \ref{small-gamma}. Moreover, observe that the correlation $C_i(t \to \infty)$ decays to zero in the Markov $(\gamma \to \infty)$ limit. However, for any finite $\gamma$, it possesses a substantial non-zero value.
In what follows, we show that the knowledge of the spin-position correlation $C_i(t)$ is essential to compute the fluctuations and correlations of the displacements of the tagged particles for active random average process.

\section{Mean squared displacement and correlations in the quenched initial condition}
\label{sec-q}
\noindent 
We now look at the mean squared displacement and the equal time correlations of the positions of the tagged particles when their initial positions are fixed as given in Eq. \eqref{quenched-eq}. However, the initial spin $\sigma _i(0)$ can still fluctuate and take values $\sigma _i(0) = \pm 1$ with equal probability $1/2$ independently for individual particles. First notice that due to the translational symmetry in our model, the correlation $\langle z_i(t) z_j(t)\rangle$ will depend only on the separation $|i-j|$. Therefore, without any loss of generality, we put $j=0$ and denote the correlation $\langle z_0(t) z_i(t)\rangle$ by $g_i(t)$. 

To derive the time evolution equation for $g_i(t)$, we follow the same procedure as done for $C_i(t)$ in Section \ref{corr-ci}. At a small time interval $[t, t+dt]$, we evaluate $g_i(t+dt)=\langle z_0(t+dt) z_i(t+dt)\rangle$ using the update rule in Eq.\eqref{update-eq-3}. Keeping all terms up to linear order in $dt$, one finds the following evolution equation for $i \neq 0$
\begin{align}
g_i(t+dt)  =  g_i(t)+\mu_1 dt \Big[g_{i+1}(t)+g_{i-1}(t)-2g_i(t)+2a C_i(t) + T_i(t)\Big], \label{two-pt-1} 
\end{align}
where $C_i(t) = \langle z_i(t) \sigma _0(t) \rangle = \langle z_0(t) \sigma _i(t) \rangle$, and $T_i(t)$ is a combination of the three-point correlations defined as 
\small{\begin{align}
	T_i(t)  = \frac{1}{2}\Big[\langle z_0(t) z_{i+1}(t) \sigma_i(t) \rangle-\langle z_{-1}(t) z_{i} (t) \sigma_0 (t) \rangle+\langle z_1(t) z_{i} (t) \sigma_0 (t) \rangle-\langle z_0 (t) z_{i-1}(t) \sigma_i (t) \rangle \Big]. \label{T-exp}
	\end{align}}
\normalsize{On} the other hand, the same procedure for  $i =0$ yields
\begin{align}
g_0(t+dt)  =  & ~g_0(t)+\mu_1 dt \Big[g_{1}(t)+g_{-1}(t)-2g_0(t)+ 2a C_0(t) + T_0(t)\Big]  \nonumber \\
& + \mu _2  dt \left[ a^2 +2 g_0(t)-2 g_1(t)+2a C_1(t)-2 a C_0(t)-T_0(t)\right].
\label{two-pt-2} 
\end{align}
Combining Eqs. \eqref{two-pt-1} and \eqref{two-pt-2} and taking $dt \to 0$ limit, we find 
\begin{align}
\frac{dg_i(t)}{dt}  = & \mu_1 \Big[g_{i+1}(t)+g_{i-1}(t)-2g_i(t)+2a C_i(t) + T_i(t)\Big] \nonumber \\
&+\delta _{i,0} ~\mu _2  \left[ a^2 +2 g_0(t)-2 g_1(t)+2a C_1(t)-2 a C_0(t)-T_0(t)\right], \label{two-pt-3} 
\end{align}
for all $i = -\infty, ...,-1,0,1,...\infty$.
For the original MRAP, the corresponding equation was derived in \cite{Rajesh2001}. Unlike in the Markov case, once again we find that Eq. \eqref{two-pt-3} does not satisfy the closure property and involves higher order correlations in the form of $T_i(t)$. Also the function $C_i(t)$ needs the knowledge of higher order correlations as illustrated in Eq. \eqref{ceq-1}. Overall this makes the computation of $g_i(t)$ for the persistent case rather challenging. To proceed, we perform the joint Fourier-Laplace transformation
\begin{align}
\bar{g}(q,t)& = \sum _{i=-\infty}^{\infty} e^{\iota i q}~ g_i(t),~~~~~~~~~~\mathcal{G}(q,s) = \int _{0} ^{\infty} dt~e^{-st}~\bar{g}(q,t), \label{FT-1}\\
\bar{C}(q,t) &= \sum _{i=-\infty}^{\infty} ~e^{\iota i q} C_i(t),~~~~~~~~~~\mathcal{C}(q,s) = \int _{0} ^{\infty} dt~e^{-st}~\bar{c}(q,t), \label{FT-2}\\
\bar{T}(q,t) &= \sum _{i=-\infty}^{\infty} e^{\iota i q}~ T_i(t),~~~~~~~~~~\mathcal{T}(q,s) = \int _{0} ^{\infty} dt~e^{-st}~\bar{T}(q,t), \label{FT-3}
\end{align}
where $\iota^2=-1$ and plug them in Eq. \eqref{two-pt-3} to obtain
\begin{align}
\mathcal{G}(q,s) &= \frac{\mu_2 }{s+\beta(q)} \left[ \frac{a^2}{s}+2\left(\tilde{g}_0(s)-\tilde{g}_1(s) \right)+2a \left(\tilde{C}_1(s)-\tilde{C}_0(s) \right) -\tilde{T}_0(s)\right] \nonumber \\
&~~~~~~~~~~~~~~~~~~~~~ +\frac{\mu_1 }{s+\beta(q)} \left[  2 a~ \mathcal{C}(q,s)+\mathcal{T} (q,s) \right] , \label{two-pt-4} 
\end{align}
where $\beta(q) = 2 \mu _1 (1-\cos(q))$ and the functions $\tilde{g}_i(s)$, $\tilde{C}_i(s)$ and $\tilde{T}_i(s)$ denote the Laplace transformations of $g_i(t)$, $C_i(t)$ and $T_i(t)$ respectively. For given $\tilde{C}_i(s)$ and $\tilde{T}_i(s)$, Eq. \eqref{two-pt-4} has two unknowns, namely $\tilde{g}_0(s)$ and $\tilde{g}_1(s)$. To get rid of $\tilde{g}_1(s)$, we take the Laplace transformation of Eq. \eqref{two-pt-3} for $i=0$ and get
\beq
\tilde{g}_0(s)-\tilde{g}_1(s)=\frac{a\mu_2 \tilde{C}_1(s) }{\mu_1-\mu_2}+a\tilde{C}_0(s)+\frac{\tilde{T}_0(s)}{2}+\frac{\mu_2 a^2}{2s (\mu_1-\mu_2)}-\frac{s~\tilde{g}_0(s)}{2(\mu_1-\mu_2)}.
\label{g0g1_relation}
\eeq 
Inserting this in Eq. \eqref{two-pt-4}, we obtain
\begin{align}
\mathcal{G}(q,s) &= \frac{\mu_2  (\mu_1-\mu_2)^{-1}}{s+\beta(q)} \left[ \frac{\mu _1a^2}{s}+2 a \mu _1 \tilde{C}_1(s)-s \tilde{g}_0(s)\right]  +\frac{\mu_1 }{s+\beta(q)} \left[  2 a \mathcal{C}(q,s)+\mathcal{T} (q,s) \right]. 
\label{two-pt-5} 
\end{align}
We now have to specify only $\tilde{g}_0(s)$ to calculate $\mathcal{G}(q,s)$. However, the Laplace transform $\tilde{g}_0(s)$ can be obtained self-consistently from Eq. \eqref{two-pt-5} by integrating $\mathcal{G}(q,s) $ with respect to $q$. Then plugging back $\tilde{g}_0(s)$ in Eq. \eqref{two-pt-5} gives the function $\mathcal{G}(q,s) $ exactly. Below, we analyse this equation first to calculate the mean squared displacement $g_0(t)$ and then the equal time correlation.

\subsection{Mean-squared displacement $g_0(t) = \langle z_0^2(t) \rangle$}
\label{msd-q}
The MSD can be obtained by taking the inverse Fourier transform $\tilde{g}_0(s) = \int _{-\pi}^{\pi}\frac{dq}{2 \pi}~ \mathcal{G}(q,s)$ with the expression of $\mathcal{G}(q,s)$ given in Eq. \eqref{two-pt-5}. We then obtain
\begin{align}
\tilde{g}_0(s) \left[ 1+\frac{\mu _2 s Y(s) }{\mu _1 - \mu _2}\right]  =&  \mu _1  W(s) + \frac{\mu _1 \mu _2 ~Y(s)}{\mu_1-\mu _2} \left[  \frac{ a^2}{s} +2 a \tilde{C}_1(s) \right],  \label{msd-eq-1} \\
\text{where}~~~~~  Y(s)  = &~ \frac{1}{2 \pi}\int _{-\pi} ^{\pi}\frac{dq}{s+\beta(q)} = \frac{1}{\sqrt{s^2+4 \mu _1  s}}, \label{func-Y}\\
\text{and}~~~~~W(s) = & ~\frac{1}{2 \pi}\int _{-\pi} ^{\pi}dq ~\frac{2 a \mathcal{C}(q,s)+\mathcal{T}(q,s)}{s+\beta (q)}. \label{func-Z}
\end{align}
This equation formally gives the exact MSD of the position of the tagged particle in the Laplace domain given the two and three point correlations $C_i(t)$ and  $T_i(t)$ are known. 
\begin{figure}[t]
	\includegraphics[scale=1.2]{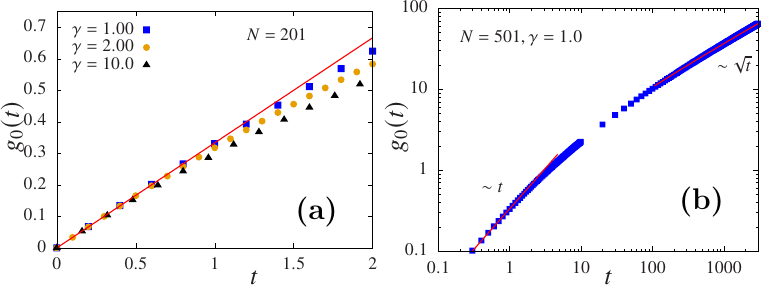}
	\centering
	\caption{(a) Comparison of the mean squared displacement $\langle z_0^2(t) \rangle $ with the numerical simulation for small values of $t$. Solid lines represent the analytical expression in Eq. \eqref{gi_small_t_theory} and symbols are from simulation. (b) Numerical verification of the crossover behaviour of $\langle z_0^2(t) \rangle $ from linear growth at small times to sub-diffusive $\sim \sqrt{t}$ growth at late times. The corresponding expressions are given in Eq. \eqref{gi_small_t_theory} for small $t$ and in Eq. \eqref{MSD-eqq-3} for large $t$. For panels (a) and (b), the MSD of a tagged particle is computed using $10^7$ and $10^6$ realizations respectively.}    
	\label{small-t-MSD}
\end{figure} 
Though the functions $\mathcal{C}(q,s)$ and $\mathcal{T}(q,s)$ are not known exactly,  one can still derive some scaling behaviors of $g_0(t)$ for different values of $t$.
For small $t$, one has $C_i(t \to 0) = 0$ and $T_i(t \to 0) = 0$ because at very small times, the displacement is negligibly small. In the Laplace domain, this implies that both $\left[ s\tilde{C}_i(s ) \right]$ and $\left[ s\tilde{T}_i(s ) \right]$ converge to zero for large values of $s$. Hence, the function $W(s)$ in Eq. \eqref{func-Z} decays faster than $\sim 1/s$ as $s$ becomes large. On the other hand, from Eq. \eqref{func-Y}, we see $ Y(s)  \simeq 1/s$ for large $s$. Using these approximations in Eq. \eqref{msd-eq-1}, we obtain for large $s$, $\tilde{g}_0 \left(s \right) ~\simeq~ \mu_2  a^2/s^2$ which in the time domain gives
\beq
g_0(t) = \langle  z_0^2(t) \rangle ~\simeq~ \mu_2  a^2 t,~~~~\text{as }t \ll 1.
\label{gi_small_t_theory}
\eeq
This linear growth of the MSD at small times has been numerically verified in Figure \eqref{small-t-MSD} (left panel). It is easy to understand the small $t$ asymptotic from the following physical reasoning. At small times, the leading order contribution to the MSD comes from those realisations where the tagged particle has made one jump while the other particles have not moved at all. The probability of observing such an event is $t$ ($pt$ in term of the unscaled time). Now the particle jumps by a random amount $\pm \eta _0 a$ depending on its spin $\sigma _0(0)$. Since $\sigma _0(0) = \pm 1$ with equal probability $1/2$, we obtain the MSD $\langle z_0^2(t) \rangle = \mu _2 a^2 \times t = \mu_2  a^2 t$ as given in Eq. \eqref{gi_small_t_theory}.

\begin{figure}[t]
	\includegraphics[scale=1.0]{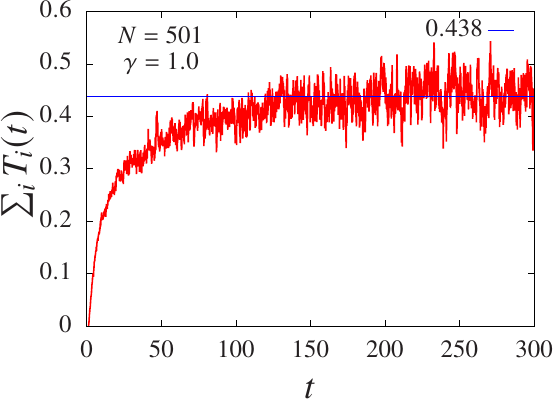}
	\centering
	\caption{Numerical plot of $\sum _{i}T_i(t)$ for $\gamma =1$, where $T_i(t)$ is defined in Eq.~\eqref{T-exp}. At late times, we find that this sum saturates to a non-zero value $ \sum _{i}T_i(t)\simeq 0.438$ which is used in Eq.~\eqref{zetaeq-1} while evaluating $g_0(t)$. We have used $10^7$ realisations to obtain the numerical data.}
	\label{fig-2-2}
\end{figure}

Next we focus on the large-$t$ behaviour of the MSD from Eq. \eqref{msd-eq-1}. Recall that for the MRAP, the MSD scales sub-diffusively as $\sim \sqrt{t}$ with a prefactor that depends on the model parameters [see Eq. \eqref{passive-case}]. To see this for the non-Markov active case, we perform the small-$s$ expansion of $ \tilde{g}_0(s)$ in Eq. \eqref{msd-eq-1} for which we need to specify $Y(s)$, $\tilde{C}_1(s )$ and $W(s)$ for smaller values of $s$.  Expression of $Y(s)$ for $s \ll 1$ follows easily from Eq. \eqref{func-Y} as $Y(s) \simeq 1/\sqrt{4 \mu _1 s}$. As mentioned earlier, due to the hierarchical dependence of the correlations it is difficult to find the small-$s$ behaviours of $\tilde{C}_1(s)$ and $W(s)$. We, however, numerically observe that the correlations $C_i(t)$ and $\sum_i T_i(t)$ at late times saturate to their $i$-dependent values and become time independent. This is numerically illustrated in Figures \eqref{fig-2} and \eqref{fig-2-2}. 
Hence for small $s$, one gets $s \tilde{C}_1(s ) \simeq C_1(t \to \infty)$ and 
\begin{align} 
W(s) \simeq \frac{Y(s)}{s}~\left[ 2 a C_I + T_I\right], ~~(\text{for } s \ll 1 )\label{ws-approx-eq}
\end{align}
where we identify 
\begin{align}
\begin{split}
\bar{C}(q \to 0, t \to \infty)=\sum_iC_i(t \to \infty)=C_I, \\ 
\bar{T}(q \to 0, t \to \infty)=\sum_iT_i(t \to \infty)=T_I
\end{split}
\label{C_I-T_I}
\end{align} 
as defined in Eq.~\eqref{def-C_I-T_I}. These constants can be obtained from the saturation values of $C_i(t)$ and $\sum_iT_i(t)$ at large $t$ [see Figures.~\eqref{fig-2} and \eqref{fig-2-2}]. Furthermore from Eq.~\eqref{func-Y}, it is easy to see that $Y(s)\simeq \frac{1}{\sqrt{4 \mu_1 s}}$ for small $s$. 
Plugging these small $s$-asymptotics in Eq. \eqref{msd-eq-1}, we find that the Laplace transform $ \tilde{g}_0(s)$ reads
\begin{align}
\tilde{g}_0(s) \simeq \frac{\sqrt{\pi} \zeta _1}{2 s^{3/2}},~~~~\text{as }s \to 0, \label{approx-gs}
\end{align} 
with the constant $\zeta _1$ given explicitly in Eq. \eqref{zetaeq-1}. {We emphasize that this expression is exact at late times and does not involve any approximation. However, it can be simplified further by using the approximate expressions of $C_i(t)$ at large $t$ given in Eq.~\eqref{ci_steady_state}.} Under this approximation, it is easy to compute $C_I=\bar{C}(q \to 0, t \to \infty)$ by taking $t\to \infty$ limit of Eq.~\eqref{ckk-eq-1} at $q=0$ and one finds 
\beq
C_I \simeq \frac{\mu_1 a }{2\gamma}, \label{C_I-approx}
\eeq
plugging which in Eq.~\eqref{zetaeq-1}, one finds the following  simpler expression for $\zeta_1$
\beq
\zeta _1 =\simeq \sqrt{\frac{\mu_1}{\pi}} \left[  \frac{a^2 \mu_1 }{\gamma}+ T_I + \frac{a^2 \mu _2 }{\mu_1-\mu_2}        \left\{ 1 + \frac{\mu_1 }{ \sqrt{ \gamma^2+\mu_1  \gamma}} ~\text{exp}{[-1/\xi]}\right\}   \right],
\label{zeta1_approximate}
\eeq
where $\xi^{-1}$ is given in Eq.\eqref{ci_steady_state-10}. It is now straightforward to perform the inverse Laplace transformation of  Eq.~\eqref{approx-gs} and obtain
\begin{align}
g_0(t) = \langle z_0 ^2(t) \rangle \simeq \zeta _1 \sqrt{t} ~~~~(\text{for large } t). \label{MSD-eqq-2}
\end{align}
\begin{figure}[]
	\includegraphics[scale=1.2]{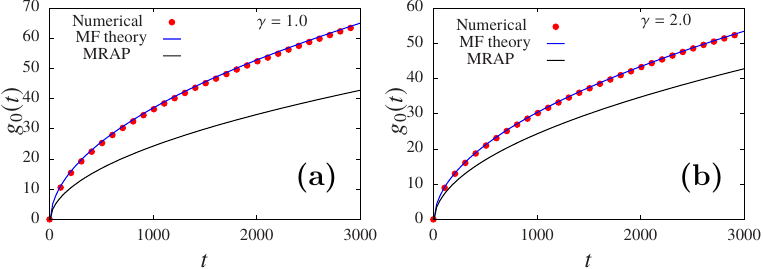}
	\centering
	\caption{Comparison of analytical (mean-field) expression of the mean squared displacement $g_0(t)=\langle z_0^2(t) \rangle $  given in Eq. \eqref{MSD-eqq-3} with the same obtained from numerical simulation for $\gamma =1$ (left panel) and $\gamma =2$ (right panel). To contrast our result, we have also plotted $\langle z_0^2(t) \rangle $ for the Markov case whose expression is given in Eq. \eqref{passive-case} [solid black line labeled by MRAP]. For comparison, we have taken $ T_I$ is $0.438$ for left panel and $0.216$ for right panel. The steady-state value of the correlator $C_i(t\to \infty)$ is approximated by the formula as given in Eq.~\eqref{ci_steady_state}. For both plots, simulation is performed with $N=501$ particles.}    
	\label{fig-MSD}
\end{figure} 
\noindent
This gives the leading order contribution to the MSD at large times. One can also obtain the sub-leading term which just turns out to be a constant. To maintain continuity of our presentation, we relegate this calculation to \ref{appen-sub-leading} and present only the final result as
\begin{align}
g_0(t) = \langle z_0 ^2(t) \rangle \simeq \zeta _1 \sqrt{t} + \zeta _2~~~~(\text{for large } t), \label{MSD-eqq-3}
\end{align}
with $\zeta_1$ and $\zeta _2$ given in Eqs. \eqref{zetaeq-1} and \eqref{zetaeq-2}, respectively. In conjunction to the Markov case, we find that the MSD for the persistent active case also scales sub-diffusively as $\sim \sqrt{t}$ at large times. Similar temporal growth patterns have been consistently observed in other active particle systems. For instance, the large-time growth of current fluctuations in run-and-tumble particles is the same as that of Brownian particles\cite{Jose2023,Tanmoyarxiv}. 
However the associated coefficients $\zeta _1$ are different for two cases. While for $\gamma \to \infty$, the two coefficients converge, we see a clear difference between them for finite $\gamma$. This difference is also illustrated in Figure \eqref{fig-MSD} where we have also compared with the numerical simulations. This implies that at large times, the tracer particle performs sub-diffusion with exponent $1/2$ for both Markov as well as the active RAP. But the persistent nature of the active particles enhances the coefficient of the MSD.  For the dynamics of a tracer particle in an active single-file system, such a difference in the MSD with respect to the Markov case was also numerically observed recently in \cite{Tirthankar}. Herein, we are able to establish this analytically based on the mean field approximations. Expectedly, this approximation breaks down for small values of $\gamma$ and one then needs to consider the exact form of $\zeta _1$. Later, we show that Eq. \eqref{approx-gs} still remains valid for small $\gamma$ and a number of results can still be derived.

To summarize, we have shown that the MSD $\langle z_0 ^2(t) \rangle $ exhibits a crossover from the diffusive scaling at small times to the sub-diffusive $(\simeq \zeta _1 \sqrt{t})$ scaling at large times with the coefficient $\zeta _1$ containing the effect of the persistent motion of active particles. This crossover behaviour has been shown in Figure \eqref{small-t-MSD} (right panel). Before we end this section, we remark that we have used the random sequential update scheme to perform the numerical simulation. For completeness, we have provided the details of this scheme in \ref{appen-simulation}. 
We use this numerical strategy to verify other analytical results throughout the paper.

\subsection{Correlation $g_i(t) = \langle z_0(t) z_i(t) \rangle$}
\label{correlation-q}
\begin{figure}[]
	\includegraphics[scale=1.2]{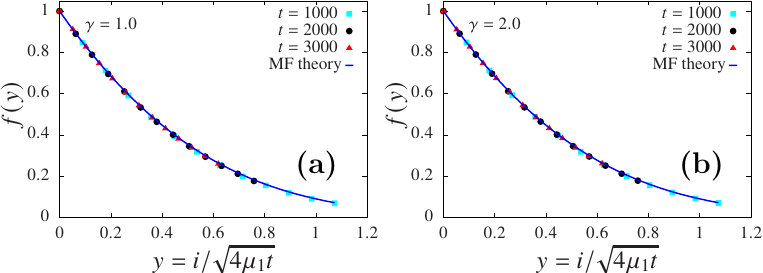}
	\centering
	\caption{Illustration of the scaling behaviour of the correlation $g_i(t) = \langle z_0(t) z_i(t) \rangle $ in Eq. \eqref{eq-corr-eq-2} for $\gamma =1$ and $\gamma =2$. Symbols are the simulation data for three different times which converge with the theoretical scaling function $f(y)$ in Eq. \eqref{scaling-func}. For both plots, simulation has been done with $N=501$ particles.}    
	\label{fig-eq-corrr}
\end{figure}

We now look at the equal time correlation of the positions of two tagged particles. For this, we have to perform the inverse Fourier transformation $\tilde{g}_i(s) = \int _{-\pi}^{\pi}\frac{dq}{2 \pi}~ e^{-i\iota q}~\mathcal{G}(q,s)$ for arbitrary $i$ and use $\mathcal{G}(q,s)$ from Eq. \eqref{two-pt-5}. We then obtain
\begin{align}
\tilde{g}_i(s)  =&  \mu _1  ~W_i(s) + \frac{ \mu _2  ~Y_i(s)}{\mu_1-\mu _2} \left[  \frac{ \mu _1 a^2}{s} +2 a \mu _1 \tilde{C}_1(s) - s \tilde{g}_0(s)\right],  \label{eq-corr-eq-1} \\
\text{where}~~~~~  Y_i(s)  = & \frac{1}{2 \pi}\int _{-\pi} ^{\pi}dq~\frac{e^{-\iota i q}}{s+\beta(q)}  \label{fun-Yr}\\
\text{and}~~~~~~~W_i(s) = & \frac{1}{2 \pi}\int _{-\pi} ^{\pi}dq~e^{-\iota i q}~\left[\frac{2 a \mathcal{C}(q,s)+\mathcal{T}(q,s)}{s+\beta (q)}\right], \label{fun-Wr}
\end{align}
with $\mathcal{C}(q,s)$ and $\mathcal{T}(q,s)$ defined in Eq.~\eqref{FT-2} and Eq.~\eqref{FT-3}, respectively.
As discussed for the MSD, carrying out the integration over $q$ in these expressions turns out to be difficult. However, for small $s$ (which corresponds to large $t$ in the time domain), one can still perform this integration approximately which then substantially simplifies the expression of $ \tilde{g}_i(s) $ in Eq. \eqref{eq-corr-eq-1}. In \ref{appen-W}, we have shown that the functions $Y_i(s)$ and $W_i(s)$ behave as
\begin{align}
& Y_i(s)   \simeq  \frac{1}{\sqrt{4 \mu _1  s}}~\exp \left[ -|i| \sqrt{\frac{s}{\mu _1 }}\right], \label{approx-Yi}\\
& W_i(s)  \simeq \frac{Y_i(s)}{s} \left[ 2 a~ C_I + T_I\right],
\label{approx-Wi}
\end{align} 
for small $s$. In addition to these quantities, we also need $\tilde{g}_0(s)$ and $\tilde{C}_1(s)$ to evaluate $ \tilde{g}_i(s)$ in Eq. \eqref{eq-corr-eq-1}. For this, we use Eq. \eqref{approx-gs} to get $\tilde{g}_0(s ) \sim s^{-3/2}$ and Eq. \eqref{ci_steady_state} to get $\tilde{C}_1(s) \simeq C_1(t \to \infty) / s$ for smaller values of $s$. Using these approximations in Eq. \eqref{eq-corr-eq-1}, we get the leading order behaviour of $ \tilde{g}_i(s)$ as
\begin{align} 
\tilde{g}_i(s) \simeq \frac{\sqrt{\pi } \zeta _1}{2 s^{3/2}}~\text{exp}\left[ -|i| \sqrt{\frac{s}{\mu _1 }}\right],~~~~~\text{for }s \ll 1. \label{approx-gi}
\end{align} 
To get the correlation $g_i(t)$, we now use the following standard Laplace transformation \cite{Rajesh2001}:
\begin{align}
\int _0 ^{\infty}dt~e^{-s t} \sqrt{\frac{t}{\pi}}  \left[ e^{-\frac{b^2}{4 t}} - \frac{b\sqrt{\pi} }{2 \sqrt{t}} ~\text{erfc} \left( \frac{b}{2 \sqrt{t}}\right)\right] = \frac{e^{-b \sqrt{s}}}{2 s^{3/2}},~~~~~\text{with }b\geq 0.
\end{align}
Plugging this in Eq. \eqref{approx-gi}, we find that $g_i(t)$ satisfies the scaling relation
\begin{align}
g_i(t) = \langle z_0(t) z_i(t) \rangle \simeq   ~\zeta _{1} \sqrt{t}~ f \left( \frac{|i|}{\sqrt{4 \mu_1  t}}\right), \label{eq-corr-eq-2}
\end{align}
with the scaling function $f(y)$ given in Eq. \eqref{scaling-func}. Once again, we observe that the correlation $g_i(t)$ is characterised by the same scaling function $f(y)$ as the MRAP in Eq. \eqref{passive-case}. However, in conjunction to the MSD, here also the signature of activity is found in the coefficient $\zeta _1$ that appears in the scaling relation. This means while the scaling function $f(y)$ is same for the two cases, the overall scaling form is slightly different for any finite $\gamma$. In Figure \eqref{fig-eq-corrr}, we have numerically illustrated this scaling behaviour for three different values of $t$ and for two different values of $\gamma$. For all cases, the simulation data converge to Eq. \eqref{scaling-func} under appropriate scaling.
\section{Mean squared displacement and correlations in the annealed initial condition}
\label{sec-a}
In the previous sections, we calculated the MSD and the correlations of the positions of tagged particles in the quenched case during which their initial positions are fixed to Eq.~\eqref{quenched-eq} for all realisations. For this case, we saw that the persistence nature of the active particles has effects on the dynamics of a tagged particle even at large times. In this section, we are interested in carrying out this analysis for the annealed case where we assume the initial positions are chosen from the stationary state of the system. Consequently, the initial positions fluctuate across different realisations. Various studies for the single-file motion of passive particles have shown that the fluctuations in initial positions have a long term effect on the dynamics of a tracer particle \cite{Tirtha2022, Krapivsky2015}. In particular, the MSD in the annealed initial condition at late times is $\sqrt{2}$ times that in the quenched initial condition. In the remaining of our paper, we address two main questions: (a) How does MSD for the persistent particles behave in the annealed case? (b) Are the MSDs for two cases related as for the non-persistent single-file systems?

In order to study the annealed case, we follow the standard technique which we briefly discuss here \cite{Krapivsky2015}. Starting from the quenched initial condition, we evolve the system up to a time $t_0$ and start measuring the position till further time $(t_0 + t)$. We then define the following two-time correlation function of two tagged particles:
\begin{align}
h_i \left(t_0, t_0+t \right) & = \langle \left[  x_i(t_0+t)-x_i(t_0) \right] \left[  x_0(t_0+t)-x_0(t_0) \right]\rangle ,\\
& = \langle \left[  z_i(t_0+t)-z_i(t_0) \right]  \left[  z_0(t_0+t)-z_0(t_0) \right]\rangle ,\\
& =  g_i(t_0 +t)+g_i(t_0) -2 S_i(t_0, t_0+t), \label{autocorrelation-1}
\end{align}
where $S_i(t_0, t_0+t) = \langle z_0(t_0) z_i(t_0+t) \rangle =  \langle z_i(t_0) z_0(t_0+t) \rangle $. The later equality can be easily proved by appropriately translating  and reflecting the index $i$. In the limit $t_0 \to \infty$, the two-time correlation function $h_i \left(t_0, t_0+t \right)$ reduces to the position correlation $l_i(t)$ in the annealed initial condition, i.e.
\begin{align}
l_i(t) &= \lim _{t_0 \to \infty} h_i \left(t_0, t_0+t \right),\\
& = \lim _{t_0 \to \infty} \left[  g_i(t_0 +t)+g_i(t_0) -2 S_i(t_0, t_0+t) \right]. \label{anneal-var-1}
\end{align}
This means that we do not need to perform the averaging over the initial positions and simply use the results for the quenched case. Since by performing the time shift by $t_0$ and taking $t_0 \to \infty$, we are effectively putting the system to its steady state, we anticipate the two methods to be equivalent. 
In what follows, we use the relation \eqref{anneal-var-1} and compute the correlation and fluctuation in the annealed setting. As clear from this relation, this reduces to calculating the correlation $S_i(t_0, t_0+t)$ which we carry out below.

For this, we again start with the evolution of $S_i(t_0, t_0+t+dt) = \langle z_0(t_0) z_i(t_0+t+dt) \rangle$ in a small time interval $[t, t+dt]$ and use Eq. \eqref{update-eq-3} to plug $z_i(t_0+t+dt)$. Following same steps as before, the dynamics of $S_i(t_0, t_0+t)$ can be shown to be
\begin{align}
\frac{\partial S_i(t_0, t_0+t) }{\partial t}  & = \frac{\mu _1 }{2} \left[ S_{i+1}(t_0, t_0+t)+S_{i-1}(t_0, t_0+t)-2 S_i (t_0, t_0+t) + 2 a C_{i}(t_0, t_0+t) \right] \nonumber \\
& + \frac{\mu _1 }{2} \left[   \langle z_0(t_0) \sigma _i(t_0+t) z_{i+1}(t_0+t)   \rangle - \langle  z_0(t_0) \sigma _i(t_0+t) z_{i-1}(t_0+t) \rangle    \right].
\label{autocorrelation-2}
\end{align}
where we have defined $C_i(t_0, t_0+t) = \langle  z_0(t_0) \sigma _i(t_0+t) \rangle$. Again, we see that this equation is not closed and involves higher order correlations. Under the mean field approximation, we break these higher correlations as a product of lower order correlations. Unlike in the equal time case, this approximation turns out to be quite good here since at large $t$, these three point correlations rapidly decay to zero. Thus within this approximation, Eq. \eqref{autocorrelation-2} can be simplified as
\begin{align}
\frac{\partial S_i(t_0, t_0+t) }{\partial t}  & \simeq \frac{\mu _1 }{2} \left[ S_{i+1}(t_0, t_0+t)+S_{i-1}(t_0, t_0+t)-2 S_i (t_0, t_0+t) + 2 a C_{i}(t_0, t_0+t) \right].
\label{autocorrelation-3}
\end{align}
\begin{figure}[t]
	\includegraphics[scale=1.2]{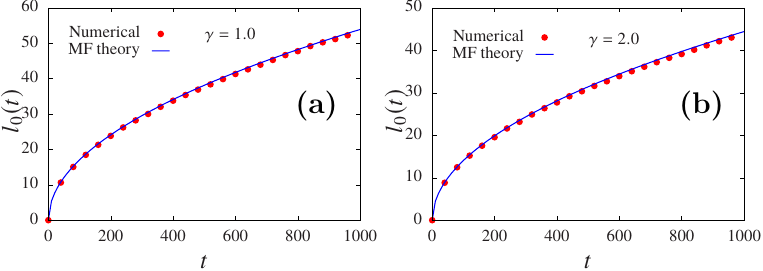}
	\centering
	\caption{Numerical verification of the MSD $l_0(t)$ with the steady state initial condition for two values of $\gamma $. The theoretical expression is given in Eq. \eqref{anneal-var-2}. For simulation, we have first allowed the system to evolve till time $t_0=3000$ and then start measuring the position with $N = 501$ particles.}    
	\label{fig-auto-corrr}
\end{figure}
\noindent
Now to solve this equation, we have to find the source term $C_i(t_0, t_0+t)$ on the right hand side. Once again this term can be easily calculated using the update rule for $\sigma _i(t)$  in Eq. \eqref{update-eq-2}. Since this derivation is exactly same as the previous ones, we have presented it in \ref{appen-Cit0t} and quote only the final result as
\begin{align}
C_i(t_0, t+t_0) = C_i(t_0)~ e^{-2 \gamma t}, \label{citot}
\end{align}
where $C_i(t_0) =\langle  z_0(t_0) \sigma _i(t_0) \rangle $ is given in Eq. \eqref{soln_ci}. We now have all terms in the right hand side of Eq. \eqref{autocorrelation-3} which can now be straightforwardly solved by taking the joint Fourier-Laplace transformation. As shown in \ref{appen-anneal-var}, we obtain the expression of $S_i(t_0,t_0+t)$ as
\begin{align}
S_i(t_0,t_0+t) \simeq \zeta_1 \sqrt{t_0}-\frac{\zeta_1 \sqrt{t}}{\sqrt{2}} ~\mathcal{W}\left( \frac{|i|}{\sqrt{2\mu_1t}} \right), \label{autocorrelation-new-5}
\end{align}
where the constant $\zeta _1$ is given in Eq. \eqref{zetaeq-1} and the function $\mathcal{W}(y)$ is defined as
\begin{align}
\mathcal{W}(y)  =   e^{-y^2}+\sqrt{\pi}y ~\rm{erf}(y).
\end{align}
We emphasize that the expression of $S_i(t_0, t_0+t) $ in Eq. \eqref{autocorrelation-new-5} is valid only for large $t_0$ and large $t$ but with their ratio $t / t_0$ fixed to a value much smaller than $1$. Below we use this form in Eq. \eqref{anneal-var-1} to compute the asymptotic behaviours of the MSD and correlation in the annealed initial setting. 
\subsection{Mean squared displacement $l_0(t)$}
\label{msd-a}
Using the result in Eq. \eqref{autocorrelation-new-5}, we get $S_0(t_0, t_0 + t ) \simeq \zeta _1 \sqrt{t_0}-\zeta_1 \sqrt{t} / \sqrt{2}$. Plugging this in Eq. \eqref{anneal-var-1} along with $g_0 (t_0 )$ from Eq. \eqref{MSD-eqq-2} for large $t_0$, we obtain the MSD $l_0(t)$ as
\begin{align}
l_0(t) \simeq \zeta _1 \sqrt{2 t}, ~~~~~~\text{for~~~} t \gg 1.
\label{anneal-var-2}
\end{align}
We have verified this analytic expression against simulations in Figure \eqref{fig-auto-corrr}. Few remarks are in order. First, the MSD of a tracer in ARAP again scales sub-diffusively as $\sim \sqrt{t}$ at large times (reminiscent of the Markov case). However, due to the persistent motion of active particles, once again we see that the coefficient accompanying this sub-diffusive growth is different than the Markov case. Only for $\gamma \to \infty$, the coefficient converges to the non-persistent value. Second interesting observation is that while MSDs in the annealed and quenched initial settings change from their Markov counterparts, their ratio is still equal to $\sqrt{2}$, a result seen for many single-file systems \cite{Tirtha2022, Krapivsky2015, Rajesh2001}. This implies that even though $l_0(t)$ and $g_0(t)$ individually carry the signature of activity, their ratio however is still fixed to the value $\sqrt{2}$ even for finite $\gamma$. 
\begin{figure}[t]
	\includegraphics[scale=1.2]{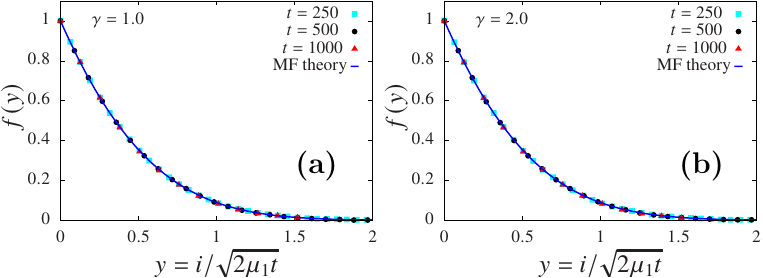}
	\centering
	\caption{Numerical verification of the scaling function $f(y)$ in Eq. \eqref{annea-corrrt-eq-1} for the correlation $l_i(t)$ in the steady state. In both panels, we have performed the comparison for three different times shown in different colours. We observe that the data for different times (symbols) converge to the theoretical curve (solid line) under scaling with respect to time. For simulation, we have first evolved the system till time $t_0=3000$ with $N=501$ and then start measuring the position.}    
	\label{fig-anneal-corrr}
\end{figure}
\subsection{Correlation $l_i(t)$}
\label{correlation-a}
We next look at the expression of $l_i(t)$ in Eq. \eqref{anneal-var-1} for general $i$ and insert $S_i(t_0,t_0+t)$ from Eq. \eqref{autocorrelation-new-5} and $g_i(t_0)$ from Eq. \eqref{eq-corr-eq-2}. The correlation then turns out to be
\begin{align}
l_i(t) & \simeq 2 \zeta_1 \sqrt{t_0} ~f \left( \frac{|i|}{\sqrt{4 \mu _1  t_0}}\right)  - 2 \zeta _1 \sqrt{t_0} +  \zeta_1 \sqrt{2t}~\mathcal{W}\left( \frac{|i|}{\sqrt{2\mu_1t}} \right), \nonumber   \\
& \simeq  - \frac{\sqrt{\pi} \zeta _1 |i| }{\sqrt{\mu _1 }} ~\text{erfc} \left( \frac{|i|}{\sqrt{4 \mu _1  t_0}} \right)
+  \zeta_1 \sqrt{2t}~\mathcal{W}\left( \frac{|i|}{\sqrt{2\mu_1t}} \right),  \nonumber  \\
& \simeq  - \frac{\sqrt{\pi} \zeta _1 |i| }{\sqrt{\mu _1 }} 
+  \zeta_1 \sqrt{2t}~\mathcal{W}\left( \frac{|i|}{\sqrt{2\mu_1t}} \right),  \nonumber  \\
& \simeq \zeta _1 \sqrt{2 t} ~f \left( \frac{|i|}{\sqrt{2 \mu _1  t}}\right) ,
\label{annea-corrrt-eq-1}
\end{align}
where the scaling function $f(y)$ is given in Eq. \eqref{scaling-func}. Also, in going from second line to the third line, we have used the asymptotic behaviour of complementary error function as $\text{erfc} \left( \frac{|i|}{\sqrt{4 \mu _1  t_0}} \right) \to 1$ as $t_0 \to \infty$ for finite $i$. In Figure \eqref{fig-anneal-corrr}, we have compared the scaling behaviour of $l_i(t)$ with the numerical simulation for $\gamma =1$ in left panel and $\gamma =2$ in right panel. For both panels, we have carried out the comparison for three different values of $t$. We observe excellent match of our analytical results with the simulation for all cases. Compared to the Markov case, once again we see that the persistence only changes the prefactor in the scaling relation \eqref{annea-corrrt-eq-1} but keeps the form of the scaling function the same.
\section{Effect of small $\gamma$ on the MSD}
\label{small-gamma}
\begin{figure}[t]
	\includegraphics[scale=1.2]{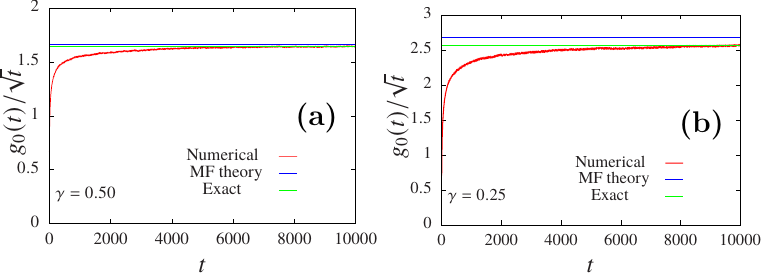}
	\centering
	\caption{Numerical verification of the prefactor $\zeta_1$ associated with the variance $g_0(t) \simeq \zeta_1 \sqrt{t}$. We have also performed a comparison with the theoretical mean-field expression in Eq.~\eqref{zeta1_approximate} (shown in blue) and the exact expression in Eq.~\eqref{zetaeq-1} (shown in green) with $C_i(t)$ obtained from simulations. Based on the simulation, we have obtained (a) $C_I= 0.485$ and $T_I= 0.91 $ (for the left panel) and (b) $C_I= 0.91$ and $T_I= 2.08 $ (for the right panel). Simulation for both plots has been conducted with $N=1001$ particles and the averaging is done over $5 \times 10^5$ realizations.}   
	\label{figure_small_gamma}
\end{figure}

In the previous sections, we looked at the fluctuations and correlations of the positions of tagged particles and studied their dependence on the initial condition. Moreover, based on a mean field approximation, we provided semi-analytic expressions of these quantities. However, it turns out that this approximation is valid only for large and intermediate values of the flipping rate $\gamma$ and breaks down for its smaller values. To illustrate this, we have plotted the simulation data of the MSD $g_0(t)/\sqrt{t}$ for $\gamma = 0.5$ and $\gamma  = 0.25$ in Figure \eqref{figure_small_gamma}. Furthermore, we compare our numerical result with its analytic form in Eq. \eqref{zetaeq-1} by computing the prefactor $\zeta _1$ in two ways: (i) first we obtain $C_i(t)$ from numerics and plug it in Eq. \eqref{zetaeq-1} to get a complete numerical estimate of $\zeta _1$, (ii) second we use the approximated theoretical form of $\zeta _1$ in Eq.~\eqref{zeta1_approximate}. As seen in Figure \eqref{figure_small_gamma}, while $\zeta_1$ for case (i) matches with the simulation data, there is a clear departure of $\zeta_1$ obtained for case (ii). This departure becomes more pertinent as we go to smaller and smaller values of $\gamma $. Hence, we still find that, $g_0(t)$ scales sub-diffusively as $\sim \sqrt{t}$ at late times even for small $\gamma$ with $\zeta _1$ given by its exact form in Eq.~\eqref{zetaeq-1}.

\begin{figure}[t]
	\includegraphics[scale=1.2]{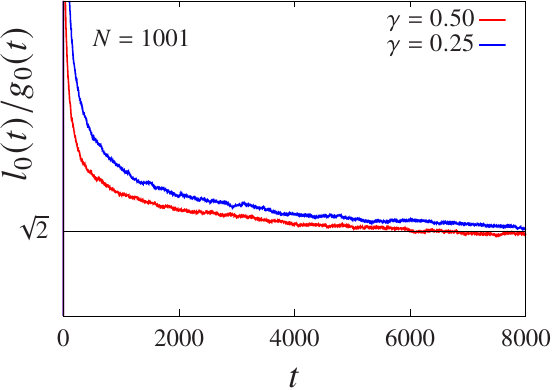}
	\centering
	\caption{Numerical comparison  of the ratio of the MSD $l_0(t)$ measured in the steady state and the MSD $g_0(t)$ with quenched uniform initial state for two small values of $\gamma$.  We observe that even for small $\gamma$, where the predictions from mean field approximation do not hold, the ratio $l_0(t) / g_0(t)$, at large $t$ approaches, $\sqrt{2} $ implying that the relation in Eq.~\eqref{eq:relation_msd_unmsd} to be  valid for all non-zero $\gamma$. For the annealed case, we start measuring position only after time $t_0=10000$ in this simulation. We have performed averaging over $5 \times 10^5$ realisations.}    
	\label{figure_small_gamma_ratio}
\end{figure}

\begin{figure}[t]
	\includegraphics[scale=1.]{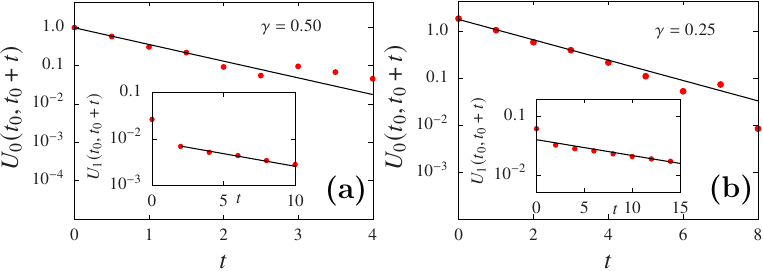}
	\centering
	\caption{Numerical plot of the correlation $U_i(t_0,t_0+t)$ in Eq.~\eqref{eq:correlation_u} for $i=0$ as a function of time for (a) $\gamma=0.5$ and (b) $\gamma=0.25$ with $t_0=10000$, $N=1001$ and $5 \times 10^5$ realisations. For large $t$, we find that the simulation data (shown in red) can be fitted (shown in black) by $\lim_{t_0 \to \infty} U_0(t_0,t_0+t)  \simeq \psi_0 ~e^{-2 \gamma  t}$ with $\psi_0=0.95$ for panel (a) and $\psi_0=1.8$ for panel (b). In the respective insets, we have shown the corresponding plots of $U_1(t_0,t_0+t)$  along with the fits $\lim_{t_0 \to \infty} U_1(t_0,t_0+t) \simeq \psi_1 ~e^{-\gamma t/4}$ with $\psi_1=0.009$ for $\gamma=0.5$ and $\psi_1=0.04$ for $\gamma=0.25$. For this case, we have chosen $t_0=200$, $N=101$ and $10^9$ realisations.
	}  
	\label{figure_decay_small_gamma}
\end{figure}

We next look at the MSD $l_0(t)$ in the annealed case where the initial positions are drawn from the steady state. For large and intermediate $\gamma$, we proved before that the ratio $l_0(t) / g _0(t)$, at late times, is still equal to $\sqrt{2}$, a relation true for many single-file systems. In the remaining part of this section, we analyze how this ratio changes for smaller values of $\gamma$. Extensive numerical simulations suggest that this ratio is still equal to the factor $\sqrt{2}$ even for small $ \gamma $. For example, in Figure \eqref{figure_small_gamma_ratio}, we have shown the simulation data for the ratio $l_0(t) / g_0(t)$ for $\gamma = 0.5$ and $\gamma = 0.25$. For both cases, we find that the ratio approaches the value $\sqrt{2}$ at late times. This means while both $l_0(t)$ and $g_0(t)$ deviate from their mean field forms, their ratio is still fixed to $\sqrt{2}$.  To prove this, we first have to evaluate the behavior of $ S_0(t_0, t_0+t) $ [see Eq. \eqref{anneal-var-1}]. Rewriting the time evolution equation for $ S_{i}(t_0, t_0+t) $ in Eq. \eqref{autocorrelation-2}, we get
\begin{align}
\frac{\partial S_i(t_0, t_0+t) }{\partial t}  & = \frac{\mu _1 }{2} \left[ S_{i+1}(t_0, t_0+t)+S_{i-1}(t_0, t_0+t)-2 S_i (t_0, t_0+t) + 2 a U_{i}(t_0, t_0+t) \right],
\label{eq:uncorrelation}
\end{align}
where the function $U_{i}(t_0, t_0+t)$ denotes
\begin{align}
U_{i}(t_0, t_0+t) = &  \frac{1}{2a} \left[   \langle z_0(t_0) \sigma _i(t_0+t) z_{i+1}(t_0+t)   \rangle - \langle  z_0(t_0) \sigma _i(t_0+t) z_{i-1}(t_0+t) \rangle    \right] \nonumber \\
&  ~~~~~~~~~~~~~~ +C_{i}(t_0, t_0+t).
\label{eq:correlation_u}
\end{align}
Due to the presence of $\sigma $-variable, we anticipate $U_{i}(t_0, t_0+t)$ to decay, at late times, as  
\begin{align}
U_{i}(t_0, t_0+t) ~~\overset{\mathrm{t_0 \to \infty}}{\simeq}~~ \psi_i ~  e^{-\chi_i t},~~~~\text{for } t \gg 1,
\label{eq:correlation_u_ansatz}
\end{align} 
where both $\psi_i$ and $\chi_i$ are time-independent functions of the index $i$. Under mean-field approximation, we explicitly showed that $\psi_i = C_i(t_0 \to \infty) $ and $\chi_i=2\gamma$ (see Eq.~\eqref{citot}). Here, we have carried out extensive numerical simulations to find out that $\chi_0= 2 \gamma$ and $\chi_1=\gamma/4$. We have illustrated  this in  Figure~\ref{figure_decay_small_gamma} for $\gamma=0.5$ and $\gamma=0.25$. Proceeding with this form, we have shown in \ref{appen-anneal-var-smallgamma} that even for small $\gamma$ we obtain
\begin{align}
S_0 \left( t_0,t_0+t \right)  \simeq \zeta _1 \sqrt{t_0} - \zeta _1 \sqrt{\frac{t}{2}},
\label{eq:unequal_autocorrelation}
\end{align}
for both $t_0 $ and $t$ large. We emphasize that this expression is valid for all non-zero values of $ \gamma $ with $\zeta _1$ given exactly in Eq. \eqref{zetaeq-1}. Plugging this in Eq. \eqref{anneal-var-1} yields
\begin{align}
l_ 0(t) \simeq \sqrt{2}~g_0(t),
\label{eq:relation_msd_unmsd}
\end{align}
with $g_0(t) = \zeta _1 \sqrt{t}$. This means that while both $l_0(t)$ and $g_0(t)$ depart individually from their mean field forms, the ratio is still given by $\sqrt{2}$. We have numerically verified this result in Figure \eqref{figure_small_gamma_ratio} for $\gamma =0.5$ and $\gamma =0.25$. Demonstrating this result numerically for very small values of $ \gamma $ turns out to be computationally expensive. Recall from Eq. \eqref{anneal-var-1} that one needs to go to very large $t_0$ to measure $l_0(t)$. Numerically, we see that smaller the value of $\gamma$, larger is the value of $t_0$ that one has to consider. On the other hand at very large time, the boundary effects start to become important which alter the MSD. In order to avoid boundary effects, we have to increase the number of particles in the simulation, which makes the computation highly expensive. In our study, we have fixed the smallest value as $\gamma =0.25$. Already for this value, we observe departure of the simulation data from mean field results.

\section{Conclusion}
\label{conclusion}
In this paper, we have investigated the motion of tracer particles in the random average process of persistent active particles in an infinite line. Using mean field approximation, we calculated the mean squared displacement and correlation of the positions of tracer particles both in the quenched initial condition and in the steady state. In particular, for the quenched case,  we showed that the MSD exhibits a crossover from a diffusive scaling at small times to a sub-diffusive $(\sim \sqrt{t})$ scaling at late times. Interestingly we find that the coefficient associated with this sub-diffusive growth is different from the corresponding non-persistent case and the two converge only in the limit $\gamma \to \infty$. For finite $\gamma$, we see a clear difference between them as illustrated in Figure \eqref{fig-MSD}. Similarly, for the position correlation, we find slight difference in Eq. \eqref{eq-corr-eq-2} compared to the Markov case. While the overall scaling function $f(y)$ in Eq. \eqref{eq-corr-eq-2} is same as the MRAP, the prefactor $\zeta _1$ is different and therefore carries the effect of the activity.

Next, we studied these quantities in the steady state where we first evolve the system till time $t_0 \to \infty$ and then start measuring the positions. Unlike in the quenched case, here the positions at the onset of the measurement fluctuate for different realisations. For this case, we analytically showed that the MSD at late times grows sub-diffusively as $\sim \sqrt{t}$ with the associated coefficient once again different than the Markov case. Only for $\gamma \to \infty$, the two become equal. Quite remarkably while both MSDs in the quenched initial condition and in the steady state individually change due to the persistent nature of the particles, their ratio is still equal to $\sqrt{2}$ at large times. This is a common result known to be true for many single-file systems \cite{Rajesh2001, Krapivsky2015, Tirtha2022}. Our study here reveals it to be valid even for the ARAP for all non-zero values of $\gamma$. Finally, we calculated the correlation between positions of two tagged particles at steady state in Eq. \eqref{annea-corrrt-eq-1}.

Most of our derivations in this paper relied on the mean-field approximation. Due to the non-Markovian nature of the noise, the equations for correlation functions do not satisfy closure property and involve higher order correlations. Moreover, we observe that these higher order correlations possess non-negligible values. To compute these in Eq.~\eqref{ceq-1} for $C_i(t) = \langle z_i(t) \sigma _0(t) \rangle $, we have used mean-field approximation. For this case, we found that the cross-correlation involved in this approximation is negligible for moderate and large tumbling rate $\gamma$. However, for the three-point correlation $T_i(t)$ in \ref{MF_four_point}, such approximation fails even for moderate $\gamma$. Only in the Markovian limit ($\gamma \to \infty$), this can be neglected. 

Solving single-file motion for active particles is a notoriously challenging problem. Here, we showcased an example of active single-file motion for which we could derive many results analytically. Carrying out this study for active particles with hardcore exclusions is an interesting and important direction to explore. Recent numerical studies in this direction have pointed out at some interesting qualitative differences than the usual single-file diffusion \cite{Tirthankar}. Proving this analytically is still an open problem. For systems obeying diffusive hydrodynamics, the coefficient of the sub-diffusive growth of the MSD of a tracer particles is specified by the diffusivity and mobility of the system \cite{Krapivsky2015}. In short range interacting systems such transport coefficients are usually determined by the two point correlations \cite{punyabrata-kundu} as in the Markov RAP case where only $\mu_1$ and $\mu_2$ appears in the expression of the MSD [see Eq.~\eqref{passive-case}]. In contrast, for our active RAP system we observe that the MSD gets contribution from higher point correlations also. It would be interesting to investigate if it is possible to derive the same MSD form as in Eq.~\eqref{eq_MSD} from  hydrodynamic evolution for the density \cite{Tanmoy2024}, more precisely for the inter-particle separation field similar to the Markov RAP case \cite{JCividini2016}. Also, in this work, we have only looked at the MSD and the two-point correlation functions. Obtaining higher moments and distribution of the position of a tagged particle are interesting problems even for the Markov RAP. Finally it would be interesting to study the effect of local biases in the single-file model of active particles in the same spirit as in \cite{Cividini2016, JCividini2016}.

\section*{Acknowledgements}
The authors thank Tirthankar Banerjee for stimulating discussions on the paper. A. K. acknowledges the support of the core research grant no. $\text{CRG}/2021/002455$ and MATRICS grant $\text{MTR}/2021/000350$ from the Science and Engineering Research Board (SERB), Department of Science and Technology, Government of India. P. S. acknowledges the support of Novo Nordisk Foundation grant $\text{NNF}21\text{OC}0071284$. The authors also acknowledge the support from the Department of Atomic Energy, Government of India, under Project No. RTI4001.

\begin{appendix}
	\section{Failure of mean field approximation to specify $T_i(t)$ in Eq.~\eqref{T-exp}}
	\label{MF_four_point}
	In this appendix, we will demonstrate why the mean field approximation does not correctly characterize the correlation $T_i(t)$ in Eq.~\eqref{T-exp}. Looking at this expression, we see that we have to calculate three-point correlation functions like $ \langle z_0(t) z_{i+1}(t) \sigma_i(t) \rangle $ and $\langle z_0(t) z_{i-1}(t) \sigma_i(t) \rangle $. To see if the mean field approach works for this case, we define a general correlation $\mathcal{T}_{ij}(t) \equiv \langle z_0(t) z_{i+1}(t) \sigma_j(t) \rangle $ and write its time evolution equations as
	\begin{align}
	\frac{d\mathcal{T} _{ij} (t)  }{dt}= & -2\gamma \mathcal{T}_{ij} (t)+\frac{\mu_1 }{2} \Big[ \mathcal{T}_{i+1,j} (t)+\mathcal{T} _{i-1,j} (t)+\mathcal{T}_{i+1,j+1}(t)+\mathcal{T}_{i-1,j-1}(t)-4 \mathcal{T}_{i,j}(t) \Big.  \nonumber \\
	& \Big.  + 2a \{\langle \sigma_0 z_{i+1}(t) \sigma_j (t) \rangle+\langle \sigma_{i+1} (t) z_0 (t) \sigma_j (t) \rangle  \Big] +                     \frac{\mu_1}{2} \Big[ \langle z_0 (t) \sigma_{i+1} (t) z_{i+2} (t) \sigma_j (t) \rangle \Big.\nonumber \\
	&   \Big.   -\langle z_0 (t) \sigma_{i+1} (t)  z_{i} (t) \sigma_j (t)\rangle 
	+\langle z_1 (t) \sigma_{0} (t) z_{i+1} (t) \sigma_j (t) \rangle-\langle z_{-1} (t) \sigma_{0}(t)  z_{i+1}  (t) \sigma_j (t) \rangle \Big] \nonumber \\
	&+\mu_2  \Big[ 2 \langle z_0 (t) z_0 (t) \sigma_j(t) \rangle-\langle z_1 (t)z_1(t) \sigma_j (t)\rangle-\langle z_{-1} (t) z_{-1} (t) \sigma_j (t) \rangle  \Big. \nonumber \\
	& \Big.  -\langle z_0 (t) z_1 (t) \sigma_0 (t) \sigma_j (t) \rangle + \langle z_0 (t) z_{-1}  (t) \sigma_0 (t) \sigma_j  (t) \rangle \Big].
	\label{three_point}
	\end{align}
	Note that we are interested in calculating $T_i(t) $ which is obtained by putting $i=j$ in $\mathcal{T} _{ij}(t)$.
	Coming to Eq. \eqref{three_point}, we observe that it does not satisfy the closure property as it contains four-point correlation functions. Once again, we use mean field approximations to break the four-point correlation in terms of lower point correlations as
	\begin{align}
	\langle z_0(t) \sigma_{i+1}(t) z_{i+2} (t) \sigma_j (t) \rangle \simeq & \langle z_0 (t) \sigma_{i+1} (t) \rangle \langle z_{i+2}(t)\sigma_j (t) \rangle +\langle z_0  (t) \sigma_{j} (t) \rangle \langle z_{i+2} (t) \sigma_{i+1} (t) \rangle   \nonumber \\
	&   ~~~~~~~~~~~~~~~  + \langle z_0 (t) z_{i+2} (t) \rangle \langle \sigma_{i+1} (t) \sigma_j  (t) \rangle. 
	\label{4-2_approximation}
	\end{align}
	\begin{figure}[t]
		\includegraphics[scale=1.2]{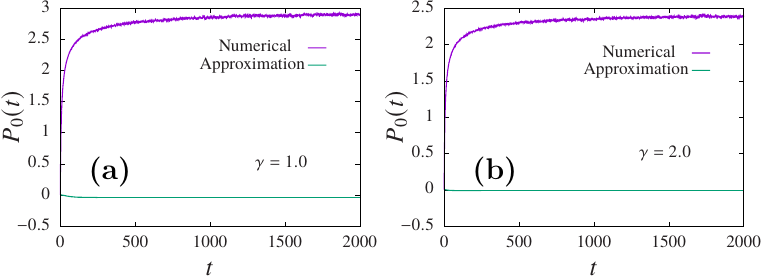}
		\centering
		\caption{Comparison of the numerically obtained four-point correlation function $P_0(t)$ as defined in Eq.~\eqref{sum_four_point}
			with its approximated value $C_1(t)(C_2(t)-C_0(t))$ using mean field. For both $\gamma =1$ and $\gamma =2$, the two deviate substantially indicating that the mean field approximation does not work for $P_0(t)$. For this simulation, we have used $4 \times 10^6$ realisations.}    
		\label{figure_MF_comparison}
	\end{figure}
	With this approximation, the following four-point point correlation function appearing in Eq. \eqref{three_point} becomes
	\begin{align}
	P_0(t)  &= \Big[ \langle z_0 (t) \sigma_{1} (t) z_{2} (t) \sigma _0 (t) \rangle -\langle z_0 (t) \sigma_{1} (t)  z_{0} (t) \sigma _0 (t) \rangle+\langle z_1 (t) \sigma_{0}(t)  z_{1}(t) \sigma_0 (t) \rangle \Big. \nonumber \\
	& ~~~~~~~~~~~~~~~~~~~~~ \Big. -\langle z_{-1} (t) \sigma_{0} (t) z_{1} (t) \sigma_0  (t)\rangle\Big], \nonumber \\
	& \simeq C_1(t) \big[C_2(t)-C_0(t)\big].
	\label{sum_four_point}
	\end{align}
	We now test the validity of this approximation. For this, we measure both $P_0(t)$ and $C_1(t) \big[C_2(t)-C_0(t)\big]$ from the numerical simulations and compare them in Figure \eqref{figure_MF_comparison} for $\gamma =1$ and $\gamma =2$. For both cases, we observe that $P_0(t)$ has a large positive value whereas $C_1(t) \big[C_2(t)-C_0(t)\big]$ has a small negative value. Clearly, this implies $P_0(t)  \neq C_1(t) \big[C_2(t)-C_0(t)\big] $. Hence we numerically find that the decoupling approximation of breaking four-point correlation function in terms of two-point correlation functions in Eq. \eqref{4-2_approximation} is not valid and thus the analytical calculation of obtaining three point correlation functions seems difficult.
	\section{Expression of $C_i(t)$ in Eq. \eqref{soln_ci}}
	\label{appen-ceq}
	In this appendix, we will derive the expression of $C_i(t)$ quoted in Eq. \eqref{soln_ci} of the main text. For this, we take the Fourier transform $\bar{C}(q,t)=\Sigma_{i=-\infty}^{\infty} e^{ \iota i q} C_i(t)$ (where ${\iota}^2=-1$) and insert this in Eq. \eqref{ceq-2} to obtain
	\begin{align}
	\frac{d\bar{C}(q,t)}{dt}=-\alpha(q) \tilde{C}(q,t)+\mu_1  a, 
	\end{align}
	where $\alpha(q)=\mu_1(1-\cos(q))+2 \gamma$. Solving this equation, we get
	\beq
	\bar{C}(q,t)=\mu_1  a \left( \frac{1-e^{-\alpha(q)t}}{\alpha(q)}\right). \label{ckk-eq-1}
	\eeq
	Performing the inverse Fourier transformation yields
	\beq
	C_i(t)=\frac{\mu_1 a}{2\pi} \int_{-\pi}^{\pi} e^{- \iota i q} \left( \frac{1-e^{-\alpha(q)t}}{\alpha(q)}\right) dq.
	\label{formal}
	\eeq
	Performing the integration over $q$ analytically in this equation is difficult. However, one can get a simplified expression by proceeding as follows. Differentiating on both sides of Eq.~\eqref{formal}, we get
	\begin{align}
	\frac{dC_i(t)}{dt}&=\frac{\mu_1 a}{2\pi}  \int_{-\pi}^{\pi} e^{- \iota i q} e^{-\alpha(q)t} dq, \notag \\
	& =\frac{\mu_1 a}{2\pi} e^{-(2\gamma +\mu_1 ) t } \int_{-\pi}^{\pi} \cos(i q) e^{\mu_1  t \cos(q)} dq, \notag \\
	& =\mu_1 a e^{-(2\gamma +\mu_1 ) t } I_{|i|}(\mu_1  t).
	\label{time_diff_eqn}
	\end{align}
	Next, we integrate both sides with respect to $t$ and use the initial condition $C_i(0)=0$ to obtain
	\beq
	C_i(t)=\mu_1 a \int_{0}^{t} e^{-(2\gamma +\mu_1 ) \tau } I_{|i|}(\mu_1 \tau) ~d \tau. 
	\label{appen-soln_ci}
	\eeq
	In the limit $t \to \infty$, one can exactly carry out the integration over $\tau$ to get
	\begin{align}
	C_i(t\to \infty)&=\mu_1 a \int_{0}^{\infty} e^{-2\gamma \tau-\mu_1  \tau } I_{|i|}(\mu_1  \tau) ~d \tau, \notag \\
	&=a \int_{0}^{\infty} e^{-\left(\frac{2\gamma +\mu_1  }{\mu_1 }\right) w} I_{|i|}(w)~dw,  \notag \\
	&=\frac{\mu_1  a}{\sqrt{4 \gamma^2+4\mu_1  \gamma}} \left(\frac{2\gamma+\mu_1 +\sqrt{4 \gamma^2+4\mu_1  \gamma}}{\mu_1 }\right)^{-|i|}.
	\label{appen-ci_steady_state}
	\end{align}
	This result has been used in Eq. \eqref{ci_steady_state} in the main text.
	\section{Sub-leading term in $g_0(t)$ in Eq. \eqref{MSD-eqq-3}}
	\label{appen-sub-leading}
	In this appendix, we derive the expression of the sub-leading term in the variance $g_0(t)$ for large $t$. As seen in Eq. \eqref{MSD-eqq-3}, for large $t$, the variance scales sub-diffusively as $g_0(t)  \simeq \zeta _1 \sqrt{t}$ with prefactor $\zeta _1$ given in Eq. \eqref{zetaeq-1}. Here, we are interested in calculating the next order correction which turns out to be a constant. To derive this, let us quote here the Laplace transform $\tilde{g}_0(s)$ from Eq. \eqref{msd-eq-1}
	\begin{align}
	\tilde{g}_0(s) \left[ 1+\frac{\mu _2 s Y(s) }{\mu _1 - \mu _2}\right]  =&  \mu _1 W(s) + \frac{\mu _1 \mu _2~Y(s)}{\mu_1-\mu _2} \left[  \frac{ a^2}{s} +2 a \tilde{C}_1(s) \right],  \label{appen-sub-leading-eq-1}
	\end{align}
	where $Y(s)$ and $W(s)$ are defined in Eqs. \eqref{func-Y} and \eqref{func-Z} respectively. Note that $Y(s ) \simeq 1 / \sqrt{4 \mu _1  s}$ for small $s$, inserting which in Eq. \eqref{appen-sub-leading-eq-1} gives
	\begin{align}
	\tilde{g}_0(s) \simeq \underbrace{\frac{\mu _1  W(s)}{1+\phi \sqrt{s}}}_{\text{first term}}      +   \underbrace{ \frac{\mu _2 \sqrt{\mu _1 } }{2\sqrt{s}~\left( \mu _1 -\mu _2 \right)  \left( 1+\phi \sqrt{s} \right)}~\left[  \frac{ a^2}{s} +2 a \tilde{C}_1(s) \right]}_{\text{second term}}, \label{appen-sub-leading-eq-2}
	\end{align}
	where $\phi = \frac{\mu _2}{\sqrt{4 \mu _1 }~(\mu _1 - \mu _2)}$. For computational ease, we have written the two terms separately in the right hand side. As evident, for the first term, we have to specify the function $W(s)$. For small $s$, the integrand in Eq. \eqref{func-Z} is dominated by smaller values of $q$. We therefore anticipate the major contribution to the integration to come from smaller values of $q$. With this approximation, the expression of $W(s)$ reduces to
	\begin{align}
	W(s) \simeq~\frac{1}{\sqrt{4 \mu _1 s}} \left[ 2 a \mathcal{C} \left(q \to 0,s \right)+\mathcal{T} \left(q \to 0,s \right)   \right],
	\end{align}
	and the first term in Eq. \eqref{appen-sub-leading-eq-2} for $s \to 0$ becomes
	\begin{align}
	\text{first term} \simeq \sqrt{\frac{\mu _1 }{4}}~\left( \frac{1}{\sqrt{s}} - \phi \right)~\left[ 2 a \mathcal{C} \left(q \to 0,s \right)+\mathcal{T} \left(q \to 0,s \right)   \right] . \label{appen-sub-leading-eq-3}\end{align}
	We now have to compute the asymptotic forms of $\mathcal{C} \left(q \to 0,s \right)$ and $\mathcal{T} \left(q \to 0,s \right)$ as $s \to 0$. To do this, we first recall that both $C_i(t)$ and $T_i(t)$ involve the $\sigma$-variables in their definitions. Due to this, both of these functions $\mathcal{C} \left(q \to 0,s \right)$ and $\mathcal{T} \left(q \to 0,s \right)$ relax exponentially to their steady values with relaxation time scale $\sim 1 / \gamma $. For $\mathcal{C} \left(q \to 0,s \right)$, one can show this from Eq. \eqref{ckk-eq-1} whereas for $\mathcal{T} \left(q \to 0,s \right)$, one can see this numerically. In the Laplace domain, this implies
	\begin{align}
	\mathcal{C}(q \to 0,s ) &\simeq~ \frac{\bar{C}(q \to 0, t \to \infty)}{s}+\text{constant term}, \\
	\mathcal{T}(q \to 0,s) &\simeq ~\frac{\bar{T}(q \to 0, t \to \infty)}{s}+\text{constant term}, 
	\end{align}
for small $s$. Here the constant terms do not involve $s$. Inserting these forms in Eq. \eqref{appen-sub-leading-eq-3} and using the identification in Eq.~\eqref{C_I-T_I}, we obtain
	\begin{align}
	\text{first term}  & \simeq \frac{\sqrt{\mu _1 }}{2}~\left[ 2 aC_I+T_I   \right] ~\left( \frac{1}{s^{3/2}} - \frac{\phi}{s} \right) , 
	\label{appen-sub-leading-eq-4_0}
	\end{align}
	which can be simplified further as 
	\begin{align}
	\text{first term}  
	& \simeq  \frac{1}{2}\left[ \frac{a^2 (\mu _1 )^{3/2}}{\gamma } + \sqrt{\mu _1 } ~T_I   \right]~\left( \frac{1}{s^{3/2}} - \frac{\phi}{s} \right), \label{appen-sub-leading-eq-4}
	\end{align}
	using the approximation $C_I=\frac{\mu_1 a}{2 \gamma}$ as derived in Eq.~\eqref{C_I-approx}. 
	So far, we have obtained the approximate form of the first term in  Eq. \eqref{appen-sub-leading-eq-2} for small $s$. Next, we carry out the same analysis for the second term. Looking at Eq. \eqref{appen-sub-leading-eq-2}, we observe that one needs to calculate the Laplace transform $\tilde{C}_1(s)$ for small $s$. To obtain this, we consider the expression of $C_1(t)$ in Eq. \eqref{soln_ci} and rewrite it as
	\beq
	C_1(t)=C_1(t \to \infty)-\mu_1 a \int_{t}^{\infty} e^{-(2\gamma +\mu_1 ) \tau }~ I_{1}(\mu_1  \tau) ~d \tau .
	\label{appen-sub-leading-eq-5}
	\eeq
	For large $t$, the argument of the Bessel function is also large which enables us to use the approximation $I_{1}(\mu_1  \tau) \simeq e^{\mu _1  \tau} / \sqrt{2 \pi \mu _1  \tau}$ for $\mu _1  \tau \gg 1$. With this approximation, the integration over $\tau$ can be easily carried out and we get 
	\beq
	C_1(t) \simeq C_1(t \to \infty)- a \sqrt{\frac{\mu _1 }{4\gamma }}~\text{erfc}\left(  \sqrt{2 \gamma t}\right).
	\label{appen-sub-leading-eq-6}
	\eeq
	Taking the Laplace transformation of this equation gives
	\begin{align}
	\tilde{C}_1(s) & \simeq \frac{C_1(t \to \infty)}{s} - a \sqrt{\frac{\mu _1 }{4\gamma }}~\frac{1}{s+2 \gamma +\sqrt{2 \gamma s + 4 \gamma ^2}}, \\
	& \simeq  \frac{C_1(t \to \infty)}{s} - \frac{a}{4 \gamma} \sqrt{\frac{\mu _1 }{4 \gamma }},~~~~\text{as }s \to 0.
	\end{align}
	Plugging this approximate form in Eq. \eqref{appen-sub-leading-eq-2}, we obtain the second term as
	\begin{align}
	\text{second term} = \frac{a \mu _2 \sqrt{\mu _1 }}{2(\mu _1-\mu _2)} \left( a + 2 C_1(t \to \infty)\right)~\left( \frac{1}{s^{3/2}} - \frac{\phi}{s} \right).
	\end{align}
	Therefore, we have computed the approximate forms of both terms in Eq. \eqref{appen-sub-leading-eq-2} for small $s$. Using these forms, the expression of the Laplace transform  $ \tilde{g}_0(s)$ simplifies to
	\begin{align}
	\tilde{g}_0(s) \simeq \frac{\sqrt{\pi}}{2}~\zeta _1 \left( \frac{1}{s^{3/2}} - \frac{\phi}{s} \right).
	\end{align}
	Finally performing the inverse Laplace transformation, we obtain the result written in Eq. \eqref{MSD-eqq-3} in the main text.
	
	\section{Details about numerical simulations}
	\label{appen-simulation}
	This appendix provides details about the numerical simulation adopted to verify various analytical results in the paper. To begin with, we have $N(=2n+1)$ number of particles initially placed at a uniform distance apart as
	\begin{align}
	x_i(0) = i a,
	\end{align}
	where $i$ is an integer that lies between $-n \leq i \leq n$. On the other hand, we choose initial $\sigma _i(0)$ from $\pm 1$ with equal probability $1/2$ independently for all particles. This means while the initial positions are fixed for different realisations, the initial $\sigma_i(0)$ still fluctuate. For a given realisation, we then implement random sequential rule to update the position of the particles. During a small time interval $[t,t+dt]$, we choose a random integer $m$ uniformly between $[-n, n]$ and update the position $x_m(t)$  and spin variable $\sigma _m(t)$ according to Eqs. \eqref{update-eq-1} and \eqref{update-eq-2} using the Gillespie algorithm~\cite{Gillespie_JCP_1976}. We then repeat this step for $N$ times. This implies that in the time interval $dt$, we perform the update rule randomly for $N$ times. Finally, we iterate this process till the observation time $t$ is reached.
	{For all figures, we chose $dt=0.01$ except for the Figures \eqref{fig-1} and \eqref{fig-2} for which we chose $dt=0.002$. The random variables $\eta_i$ in Eq.~\eqref{eq:eta} are chosen uniformly from $[0,1)$, hence $R(\eta)=1$ for $0 \le \eta <1$ and zero otherwise. To obtain the numerical data for Figs.~\ref{fig-MSD}-\ref{fig-anneal-corrr}, we have used $10^6$ realisations.}

	\section{Expressions of $Y_i(s)$ and $W_i(s)$ as $s \to 0$}
	\label{appen-W}
	Here, we derive the approximate expressions of $Y_i(s)$ and $W_i(s)$ in Eqs. \eqref{fun-Yr} and \eqref{fun-Wr} for small values of $s$. Let us first present the calculation for $Y_i(s)$ which reads
	\begin{align}
	Y_i(s)  = & \frac{1}{2 \pi}\int _{-\pi} ^{\pi}dq~\frac{e^{ -\iota i q}}{s+\beta(q)}, \label{appen-func-Y}
	\end{align}
	where $\beta(q) = 2 \mu _1 (1-\cos(q))$. For $s \to 0$, the integrand in Eq. \eqref{appen-func-Y} diverges as $q \to 0$. Therefore, we expect the major contribution to the integration should come from the small values of $q$. Taking the $q \to 0$ limit, we get $\beta (q) \simeq \mu _1 q^2$ plugging which in Eq. \eqref{appen-func-Y}, we get
	\begin{align}
	Y_i(s)  \simeq & \frac{1}{2 \pi}\int _{-\pi} ^{\pi}dq~\frac{e^{ -\iota i q}}{s+ \mu _1  q^2}. \label{appen-func-Y-2}
	\end{align}
	Changing the variable $ q = \sqrt{s/\mu _1 }~ w$ and taking $s \to 0$, we get
	\begin{align}
	Y_i(s)  & \simeq  \frac{1}{2 \pi \sqrt{s \mu _1 }}~\int _{-\pi} ^{\pi}\frac{dw}{1+ w^2}~\text{exp}\left[ - \iota i w \sqrt{\frac{s}{\mu _1 }}\right], \\
	&\simeq \frac{1}{\sqrt{4 \mu _1 s}}~\text{exp}\left[ -|i| \sqrt{\frac{s}{\mu _1 }}\right].
	\label{appen-func-Y-3}
	\end{align}
	This result has been quoted in Eq.~\eqref{approx-Yi} which was instrumental in obtaining the asymptotic behaviour of the MSD and correlation for the positions of the particles. 
	
	It turns out that for $W_i(s)$ also, one can proceed similarly to get its small $s$ behaviour. To see this, let us first rewrite its expression from Eq. \eqref{fun-Wr}
	\begin{align}
	W_i(s) = & \frac{1}{2 \pi}\int _{-\pi} ^{\pi}dq ~e^{-\iota i q}~\left[\frac{2 a \mathcal{C}(q,s)+\mathcal{T}(q,s)}{s+\beta (q)} \right],  \label{appen-func-Y-4}
	\end{align}
	where $\mathcal{C}(q,s)$ and $\mathcal{T}(q,s)$ are the joint Fourier-Laplace transforms given in Eqs. \eqref{FT-2} and \eqref{FT-3} respectively. As discussed before for small $s$, the integration in Eq. \eqref{appen-func-Y-4} will be dominated by small values of $q$ which gives
	\begin{align}
	W_i(s) \simeq & \frac{1}{2 \pi}\int _{-\pi} ^{\pi}dq ~e^{-\iota i q}~\left[\frac{2 a \mathcal{C}(q \to 0,s \to 0)+\mathcal{T}(q \to 0,s \to 0)}{s+ \mu _1 q^2} \right].  \label{appen-func-Y-5}
	\end{align}
	Numerically, we see that both $ \bar{C}(q,t) $ and $\bar{T}(q,t )$ in Eqs. \eqref{FT-2} and \eqref{FT-3} attain stationary values as $t \to \infty$. This implies, as mentioned previously,  that in the Laplace domain, one gets
	\begin{align}
	\mathcal{C}(q \to 0,s \to 0) &\simeq~ \frac{\bar{C}(q \to 0, t \to \infty)}{s},  
	\\ 
	\mathcal{T}(q \to 0,s \to 0) &\simeq ~\frac{\bar{T}(q \to 0, t \to \infty)}{s},
	\label{appen-func-Y-6}
	\end{align}
	Plugging these forms in Eq. \eqref{appen-func-Y-5} gives
	\begin{align}
	W_i(s) \simeq &~ \frac{Y_i(s)}{s}~\left[ 2 a C_I + T_I\right].
	\label{appen-func-Y-7}
	\end{align}
	where we have used Eq.~\eqref{C_I-T_I} and $Y_i(s)$ is given in Eq. \eqref{appen-func-Y-2}. Inserting the form of $Y_i(s)$ for small $s$ from Eq.~\eqref{appen-func-Y-3}, we finally get
	\begin{align}
	W_i(s) \simeq \frac{Y(s)}{s}~\left[  2a C_I+ T_I  \right]~~\text{exp}\left[ -|i| \sqrt{\frac{s}{\mu _1 }}\right].\label{appen-func-Y-8}
	\end{align}
	To summarize, in this appendix, we have derived the forms of $Y_i(s)$ and $W_i(s)$ as $s \to 0$. Observe that for $i = 0$, the function $Y_i(s)$ from Eq.~\eqref{appen-func-Y} reduces to $Y(s)$ in Eq. \eqref{func-Y}. Consequently, the Eq.~\eqref{appen-func-Y-8} with $i=0$ provides the  small $s$ behaviour of $W(s)$ as quoted in Eq. \eqref{ws-approx-eq}.
	\section{Computation of $C_i(t_0, t_0+t) = \langle  z_0(t_0) \sigma _i(t_0+t)   \rangle$}
	\label{appen-Cit0t}
	This appendix presents a derivation of the expression of $C_i(t_0, t_0+t) $ quoted in Eq. \eqref{citot}. Let us begin by writing the dynamics of $C_i(t_0, t_0+t+dt)$ in small time interval $dt$. Using the time evolution of $\sigma _i(t)$ in Eq. \eqref{update-eq-2}, we have
	\begin{align}
	C_i(t_0, t_0+t+dt) & = \langle  z_0(t_0) \sigma _i(t_0+t+dt)   \rangle, \\
	& = -\gamma dt \langle  z_0(t_0) \sigma _i(t_0+t)   \rangle +(1-\gamma dt) \langle  z_0(t_0) \sigma _i(t_0+t)   \rangle, \\
	& =  C_i(t_0, t_0+t) - 2 \gamma dt~C_i(t_0, t_0+t).
	\end{align}
	Taking the $dt \to 0$ limit, we get the dynamics of $C_i(t_0, t_0+t) $ as
	\begin{align}
	\frac{\partial C_i(t_0, t_0+t) }{\partial t} = -2 \gamma C_i(t_0, t_0+t). \label{appen-Cit0t-eq-1}
	\end{align}
	In order to solve this equation, we need to specify appropriate initial condition. Recall that as $t \to 0$, one has $C_i(t_0, t_0) = \langle z_0(t_0) \sigma _i(t_0) \rangle $ which is simply $C_i(t_0)$ defined in Section \ref{corr-ci}. The solution of $C_i(t_0)$ has been obtained in Eq. \eqref{soln_ci}. Solving Eq. \eqref{appen-Cit0t-eq-1} with this initial condition, we obtain
	\begin{align}
	C_i(t_0, t+t_0) = C_i(t_0)~ e^{-2 \gamma t},
	\end{align}
	which has also been quoted in Eq. \eqref{citot}.

	\section{Computation of $S_i(t_0, t_0+t)$ in Eq. \eqref{autocorrelation-new-5}}
	\label{appen-anneal-var}
	In this appendix, we derive the expression of $S_i(t_0, t_0+t)$ in Eq. \eqref{autocorrelation-new-5} which was used to calculate the MSD and correlation for the active RAP with annealed initial condition. To this aim, we take the joint Fourier-Laplace transformation of $S_i(t_0, t_0+t)$ as
	\begin{align}
	\mathcal{S}(q,s,t) = \sum _{i=-\infty}^{\infty} e^{ \iota i q} ~\tilde{\mathcal{S}}_i(s,t),~~~\text{with }\tilde{\mathcal{S}}_i(s,t)=\int _0^{\infty}dt_0 ~e^{-s t_0}~S_i(t_0, t_0+t), 
	\label{eq:fourier_laplace}
	\end{align}
	and insert it in Eq. \eqref{autocorrelation-3} to obtain
	\begin{align}
	\mathcal{S}(q,s,t) = \mathcal{G}(q,s) e^{-\frac{\beta(q)t}{2}} + \frac{\mu _1  a~\mathcal{C}(q,s)}{\left( \frac{ \beta(q)}{2}-2 \gamma \right)} \left( e^{-2 \gamma t}- e^{-\frac{ \beta(q)}{2}t}\right), \label{autocorrelation-4}
	\end{align}
	where $\mathcal{G}(q,s)$ and $\mathcal{C}(q,s)$ denote, respectively, the joint Fourier-Laplace transformations of $g_i(t_0)$ and $C_i(t_0)$ given in Eqs. \eqref{two-pt-5} and \eqref{ckk-eq-1}. Also, we have defined $\beta(q) = 2 \mu _1 (1-\cos(q))$. Now to get $S_i(t_0, t_0+t)$ from Eq. \eqref{autocorrelation-4}, one needs to perform two inversions: one is the inverse Fourier transformation with respect to $q$ and the other is the inverse Laplace transformation with $s$. Let us first write the inversion with respect to $q$ as
\begin{align}
    \tilde{\mathcal{S}}_i(s,t) = \int_{-\pi}^{\pi} \frac{dq}{2 \pi}~e^{ - \iota i q}~\mathcal{S}(q,s,t) = \tilde{\mathcal{S}}^{(1)}_i(s,t) + \tilde{\mathcal{S}}^{(2)}_i(s,t), \label{psnec}
\end{align}
 where the two terms are
 \begin{align}
     \begin{rcases}
         \tilde{\mathcal{S}}^{(1)}_i(s,t) &= \int_{-\pi}^{\pi} \frac{dq}{2 \pi}~e^{ - \iota i q}~ e^{-\frac{\beta(q)t}{2}}~\tilde{\mathcal{G}}(q,s)\\
         \tilde{\mathcal{S}}^{(2)}_i(s,t)&=\frac{\mu _1 a~}{2\pi} \int_{-\pi}^{\pi} \frac{dq~e^{ - \iota i q}}{\left[  \frac{ \beta(q)}{2}-2 \gamma \right]} ~       ~\mathcal{C}(q,s)\left( e^{-2 \gamma t}- e^{-\frac{ \beta(q)}{2}t}\right).
     \end{rcases}
     \label{appen-autocorrelation-eq-1}
 \end{align}
	We now proceed to evaluate these two terms separately.
	
	\subsection{Calculation of $\tilde{\mathcal{S}}^{(1)}_i(s,t)$}
	\label{first_term_calculation}
	Since we are interested in the $t_0 \to \infty$ limit, we evaluate these terms for small values of~$s$. Using the approximation  $\tilde{\mathcal{G}}(q,s) \simeq \zeta _1 \sqrt{\mu_1 \pi} /  s(s+\mu _1 q^2)$ for $s \to 0$ from Eq. \eqref{approx-gi}, the first term in Eq. \eqref{appen-autocorrelation-eq-1} becomes
	\begin{align}
	\tilde{\mathcal{S}}^{(1)}_i(s,t) & \simeq \frac{\sqrt{\mu_1 } ~\zeta_1}{2\sqrt{\pi}s }  \int_{-\pi}^{\pi} \frac{dq}{\left( s+\mu _1 q^2 \right)} ~~ \text{exp}{\Big[-\mu_1t(1-\cos(q))-i \iota q \Big]}.
	\label{unequal_smalls}
	\end{align}
	Since the integrand is exponentially decaying in $t$, the leading contribution to the integral comes from the smaller values of $q$ which enables us to approximate it as 
	\beq
	\tilde{\mathcal{S}}^{(1)}_i(s,t) \simeq \frac{\sqrt{\mu_1  } ~\zeta_1}{2\sqrt{\pi}s }  \int_{-\pi}^{\pi} \frac{dq}{\left( s+\mu _1 q^2 \right)} ~~ \text{exp}{\Big[-\frac{\mu_1 t}{2}q^2- \iota i q \Big]}.
	\eeq
	Under the variable transformation $q = \sqrt{\frac{s}{\mu_1}}w$, it simplifies to 
	\begin{align}
	\tilde{\mathcal{S}}^{(1)}_i(s,t) & \simeq  \frac{ \zeta_1}{2 \sqrt{ \pi} s^{3/2}} \int_{-\infty}^{\infty} dw ~~ \frac{\text{exp}{\Big[-\frac{st w^2}{2}- \iota i \sqrt{\frac{s}{\mu_1 }}w}\Big]}{1+w^2}, \\
	& \simeq  \frac{\sqrt{ \pi} \zeta_1}{4s^{3/2}} \left[    e^{\frac{st}{2}-|i| \sqrt{\frac{s}{\mu _1 }}} ~\text{erfc} \left( \sqrt{\frac{st}{2}} -\frac{|i|}{\sqrt{2 \mu _1 t}}  \right)   + e^{\frac{st}{2}+|i| \sqrt{\frac{s}{\mu _1 }}}  \right.   \nonumber \\
	&  \left.~~~~~~~~~~~~~~~~~~~~~~\times  ~\text{erfc} \left( \sqrt{\frac{st}{2}} +\frac{|i|}{\sqrt{2 \mu _1  t}}  \right) \right].    
	\end{align}
	Once again we take the $s \to 0$ limit to recast this equation as
	\beq
	\tilde{\mathcal{S}}^{(1)}_i(s,t) \simeq \frac{\sqrt{ \pi} \zeta_1}{2} \Big[ \frac{1}{s^{3/2}}-\frac{1}{s} \Big\{ \frac{\sqrt{2 t}}{\sqrt{\pi}} e^{-\frac{i^2}{2\mu_1  t}}+\frac{i}{\sqrt{\mu_1}} {\rm{erf}}\left(\frac{i}{\sqrt{2\mu_1  t}} \right) \Big\} \Big].\label{appen-autocorrelation-eq-2}
	\eeq 
	Now it is straightforward to perform the inverse Laplace transformation with respect to $s$. However, before that, we calculate the second term in Eq. \eqref{appen-autocorrelation-eq-1}.
	\subsection{Calculation of $\tilde{\mathcal{S}}^{(2)}_i(s,t)$}
	Observe that the second term in Eq. \eqref{appen-autocorrelation-eq-1} depends on the function $\mathcal{C}(q,s)$ whose expression is given in Eq. \eqref{ckk-eq-1} in the time domain. For small $s$, this function simplifies to 
	\begin{align}
	\mathcal{C}(q,s) \simeq \frac{\mu _1  a}{s \left[ \mu _1  (1-\cos(q))+ 2 \gamma \right]}
	\end{align}
	Plugging this in Eq. \eqref{appen-autocorrelation-eq-1} gives
	\begin{align}
	\tilde{\mathcal{S}}^{(2)}_i(s,t) \simeq \frac{(\mu _1 a)^2}{2\pi s} \int_{-\pi}^{\pi} \frac{dq~e^{ - \iota i q}}{\left[  \frac{ \beta(q)}{2}-2 \gamma \right] \left[  \frac{ \beta(q)}{2}+2 \gamma \right]} ~ \left( e^{-2 \gamma t}- e^{-\frac{ \beta(q)}{2}t}\right).
	\end{align}
	Performing integration over $q$ at this stage turns out to be difficult. However, we can carry out this by taking an additional Laplace transform with respect to $t\to \lambda $ under which the above expression becomes
	\begin{align}
	\mathbb{S}_i^{(2)}(s, \lambda) & \simeq   \frac{(\mu _1 a)^2}{2\pi s(\lambda + 2 \gamma)} \int_{-\pi}^{\pi} \frac{dq~e^{ -\iota i q}}{\left[  2 \gamma + \mu _1 (1-\cos(q)) \right] \left[  \lambda + \mu _1 (1-\cos(q)) \right]}, \\
	& \simeq \frac{(\mu _1 a)^2}{8 \pi  s \gamma ^2} ~\int _{- \pi}^{\pi}~dq~\frac{e^{-\iota i q}}{\lambda + \mu _1 q^2 /2}\label{appen-autocorrelation-eq-3},
	\end{align}
where we have used the notation $\mathbb{S}_i^{(2)}(s, \lambda)$ to denote the Laplace transform of $\tilde{\mathcal{S}}^{(2)}_i(s,t)$. Moreover, in going to the second line, we have used the approximation that for small $\lambda$ (which corresponds to large $t$), the integral is dominated by small values of $q$ which then allows us to write $1- \cos(q) \simeq q^2 /2 $. Next, we change the variable $q =  \sqrt{\frac{2 \lambda }{\mu _1 }}~w$ and take $\lambda \to 0$ to rewrite Eq. \eqref{appen-autocorrelation-eq-3} as
	\begin{align}
	\mathbb{S}_i^{(2)}(s, \lambda) & \simeq \frac{a^2~\mu_1^{3/2}}{4 \pi  s \gamma ^2 \sqrt{2 \lambda}}~\int _{- \infty}^{\infty} dw~\frac{e^{-\iota i w \sqrt{\frac{2 \lambda}{\mu _1 }}}}{1+w^2}, \\
	& \simeq  \frac{a^2(\mu_1)^{3/2}}{4 s \gamma ^2 \sqrt{2 \lambda}}e^{-|i| \sqrt{\frac{2 \lambda}{\mu _1}}}.
	\end{align}
	Finally performing the inverse Laplace transformation from $\lambda \to t$ yields
	\begin{align}
	\tilde{\mathcal{S}}^{(2)}_i(s,t) & \simeq   \frac{a^2~\mu_1 ^{3/2}}{4 s \gamma ^2 \sqrt{2 \pi t}}~\text{exp}\left(-|i|^2 / 2 \mu _1 t \right). \label{appen-autocorrelation-eq-4}
	\end{align}
	\subsection{$\tilde{\mathcal{S}}_i(s,t)$ in Eq. \eqref{psnec}}
	Comparing the two terms in Eqs. \eqref{appen-autocorrelation-eq-2} and \eqref{appen-autocorrelation-eq-4} respectively, we find that the leading order contribution to $\tilde{\mathcal{S}}_i(s,t)$ in Eq. \eqref{psnec} at large $t$ comes only from the first term. This allows us to write 
	\begin{align}
	\tilde{\mathcal{S}}_i(s,t)  \simeq \frac{\sqrt{ \pi} \zeta_1}{2} \Big[ \frac{1}{s^{3/2}}-\frac{1}{s} \Big\{ \frac{\sqrt{2 t}}{\sqrt{\pi}} e^{-\frac{i^2}{2\mu_1 t}}+\frac{i}{\sqrt{\mu_1}} \rm{erf}\left(\frac{i}{\sqrt{2\mu_1 t}} \right) \Big\} \Big].  \label{appen-autocorrelation-eq-5}
	\end{align}
	Notice that our result is valid only for large $t_0$ and large $t$ but with the ratio $t /t_0 \ll 1$. Finally taking the inverse Laplace transformation, we obtain the expression of $S_i(t_0, t_0+t)$ presented in Eq. \eqref{autocorrelation-new-5}.

	\section{Derivation of $S_0 \left( t_0,t_0+t \right)$ in Eq. \eqref{eq:unequal_autocorrelation}}
	\label{appen-anneal-var-smallgamma}
	In this appendix, we derive the expression of $S_0 \left( t_0,t_0+t \right)$ in Eq. \eqref{eq:unequal_autocorrelation} which is instrumental in deriving the form of $l_0(t)$ in Eq. \eqref{eq:relation_msd_unmsd} for small $\gamma$. To this end, we take the joint Fourier-Laplace transformation
	of Eq.~\eqref{eq:uncorrelation} with respect to $i ~(\to q)$ and $t_0~(\to s)$ to obtain
	\beq
	\frac{\partial \tilde{\mathcal{S}}(q,s,t)}{\partial t}=-\frac{\beta(q)}{2} \tilde{\mathcal{S}}(q,s,t)+\mu_1 a ~\tilde{\mathcal{U}}(q,s,t),
	\eeq
	where $\tilde{\mathcal{S}}(q,s,t)$ and $\tilde{\mathcal{U}}(q,s,t)$ are the joint Fourier-Laplace transformations of $S_i(t_0,t_0+t)$ and $U_i(t_0,t_0+t)$ respectively and $\beta (q) = 2 \mu _1 (1-\cos q)$. Solving this equation with the initial condition $\mathcal{S}(q,s,t=0) = \mathcal{G}(q,s)$, we obtain
	\begin{align}
	\tilde{\mathcal{S}}(q,s,t) = \mathcal{G}(q,s) e^{-\frac{\beta(q)t}{2}} +  \mu _1 a \int _{0}^{t} d \tau ~\exp\left[-\frac{\beta(q)}{2}(t-\tau) \right] ~\tilde{\mathcal{U}}(q,s,\tau). 
	\end{align}
	We now carry out the inversion of this expression with respect to $q$ to obtain
	\begin{align}
	\mathcal{S}_0(s,t) = \underbrace{\int _{- \pi}^{\pi} \frac{dq}{2 \pi}~e^{-\frac{\beta(q)t}{2}}~\mathcal{G}(q,s)}_{\text{first term}} +\underbrace{\mu _1 a  \int _{- \pi}^{\pi} \frac{dq}{2 \pi}~\int _{0}^{t} d \tau ~\exp\left[-\frac{\beta(q)}{2}(t-\tau) \right] ~\tilde{\mathcal{U}}(q,s,\tau)}_{\text{second term}}. \label{saikat_eq_3}
	\end{align}
	Now the first term is exactly same as Eq. \eqref{appen-autocorrelation-eq-1} and has been explicitly calculated in  \ref{appen-anneal-var} to be
	\begin{align}
	\text{first term} = \frac{\sqrt{\pi} \zeta _1}{2} \left( \frac{1}{s^{3/2}} - \frac{1}{s}\sqrt{\frac{2t}{ \pi}}\right).
	\end{align}
	On the other hand, to calculate second term, we further take a Laplace transformation with respect to $t ~(\to \lambda)$ to yield
	\begin{align}
	\text{second term} = \mu _1 a  \int _{- \pi}^{\pi} \frac{dq}{2 \pi}~\frac{\tilde{\mathcal{U}}(q,s,\lambda)}{\lambda + \mu _1(1-\cos q)}. \label{saikat-eq-1}
	\end{align}
	Once again, performing integration over $q$ turns out to be difficult since we do not know the exact form of $\tilde{\mathcal{U}}(q,s,\lambda)$. However, for small $\lambda$, the integral over $q$ will be dominated by small values of $q$ and one can then use the asymptotic form of of $\tilde{\mathcal{U}}(q,s,\lambda)$ in Eq. \eqref{eq:correlation_u_ansatz}. To see this, we approximate $1- \cos q \simeq q^2 /2$ and rewrite \eqref{saikat-eq-1} for small $\lambda$ as
	\begin{align}
	\text{second term} & \simeq \mu _1 a~ \tilde{\mathcal{U}}(q \to 0,s \to 0,\lambda \to 0)  \int _{- \pi}^{\pi} \frac{dq}{2 \pi}~\frac{1}{\lambda + \mu _1 q^2 / 2}, \\
	& \simeq a \sqrt{\frac{\mu _1}{2 \lambda }} ~\tilde{\mathcal{U}}(q \to 0,s \to 0,\lambda \to 0) .
	\label{saikat-eq-2}
	\end{align}
	Using the ansatz in Eq. \eqref{eq:correlation_u_ansatz}, one can show that $\tilde{\mathcal{U}}(q \to 0,s \to 0,\lambda \to 0) \sim~ 1 / s$ where the prefactor does not depend on $\lambda$. This means that the second term diverges as $\sim 1 /\sqrt{\lambda}$ as $\lambda \to 0$ which in the time domain implies a decay of the form $\sim t^{-1/2}$. Therefore, for large $t$, we obtain sub-leading contribution of the second term compared to the first term in Eq. \eqref{saikat_eq_3} and $\mathcal{S}_0(s,t)$ becomes
	\begin{align}
	\mathcal{S}_0(s,t) \simeq \frac{\sqrt{\pi} \zeta _1}{2} \left( \frac{1}{s^{3/2}} - \frac{1}{s}\sqrt{\frac{2t}{ \pi}}\right).
	\end{align}
	Taking the inverse Laplace transformation, we obtain the expression of $S_0 \left( t_0,t_0+t \right)$ which has been quoted in Eq. \eqref{eq:unequal_autocorrelation} in the main text.

\end{appendix}


\end{document}